\documentclass[aps, prl, amsmath, amssymb, preprintnumbers, twocolumn, showpacs, superscriptaddress]{revtex4-1}
\usepackage{amssymb, amsthm}
\usepackage{color}
\usepackage{graphicx}
\usepackage{amssymb}
\usepackage{ulem}
\usepackage{hyperref}
\usepackage{amsmath}
\usepackage{ulem} 

\newcommand{\ket}[1]{| #1 \rangle}
\newcommand{\bra}[1]{\langle #1 |}

\definecolor{grey}{rgb}{.6,.6,.6}

\definecolor{orange}{rgb}{1,0.5,0}

\newcommand{\mytitle}{Non-topological parafermions in a one-dimensional fermionic model with even multiplet pairing}

\begin{document}

\title{\mytitle}

\author{Leonardo Mazza}
\affiliation{LPTMS, CNRS, Universit\'e Paris-Sud, Universit\'e Paris-Saclay, 15 rue Georges Cl\'emenceau, 91405 Orsay, France}
\affiliation{D\'epartement de Physique, \'Ecole Normale Sup\'erieure / PSL Research University, CNRS, 24 rue Lhomond, 75005 Paris, France}

\author{Fernando Iemini}
\affiliation{ICTP, Strada Costiera 11, I-34151 Trieste, Italy}
\affiliation{Instituto de F\'isica, Universidade Federal Fluminense, 24210-346 Niter\'oi, Brazil}

\author{Marcello Dalmonte}
\affiliation{ICTP, Strada Costiera 11, I-34151 Trieste, Italy}

\author{Christophe Mora}
\affiliation{Laboratoire  Pierre  Aigrain,
\'Ecole  Normale  Sup\'erieure / PSL  Research  University,
CNRS,  Universit\'e  Pierre  et  Marie  Curie-Sorbonne  Universit\'es,
Universit\'e  Paris  Diderot-Sorbonne  Paris  Cit\'e,  24  rue  Lhomond,  75231  Paris  Cedex  05,  France}

\begin{abstract}
We discuss a one-dimensional fermionic model with a generalized $\mathbb{Z}_{N}$ even multiplet pairing extending Kitaev $\mathbb{Z}_{2}$ chain. The system shares many features with  models believed to host localized edge parafermions, the most prominent being a similar bosonized Hamiltonian and a $\mathbb{Z}_{N}$ symmetry enforcing an $N$-fold degenerate ground state robust to certain disorder. Interestingly, we show that the system supports a pair of parafermions but they are non-local instead of being boundary operators. As a result, the degeneracy of the ground state is only partly topological and coexists with spontaneous symmetry breaking by a (two-particle) pairing field. Each symmetry-breaking sector is shown to possess a pair of Majorana edge modes encoding the topological twofold degeneracy. Surrounded by two band insulators, the model exhibits for $N=4$ the dual of an $8 \pi$ fractional Josephson effect highlighting the presence of parafermions.
\end{abstract}
\maketitle

\textit{Introduction ---}
The prospect of localizing Majorana fermions at the two edges of a one-dimensional topological 
superconductor~\cite{Kitaev2001,alicea2012,Beenakker2013,Guo2016,lutchyn2017realizing} has spurred an intense theoretical and experimental activity focused on semiconductor
nanowires~\cite{wire1,wire2,mourik2012,Albrecht2016,Deng1557,Chene1701476} and chains of magnetic adatoms~\cite{choy2011,NadjPerge2013,Braunecker2013,*Vazifeh2013,*klinovaja2013,yazdani2014,ruby2015,pawlak2016,feldman2017} coupled to conventional superconductors.
Not only do the Majorana fermions reveal the topological nature of the bulk groundstate, but their
non-Abelian statistics also holds promises for realizing topologically protected qubit gates
and quantum computation~\cite{nayak2008,kitaev2003,aasen2016}.
Parafermions are $\mathbb{Z}_{N}$ fractional generalizations of $\mathbb Z_2$ Majorana fermions~\cite{fendley2012,PhysRevB.87.195422,alicea2016,fendley2014,cobanera2014,PhysRevB.90.195101,mong2014,jermyn2014,zhuang2015,stoudenmire2015,Sreejith2016,Alexandradinata2016,chen2016,hiromi2017,meidan2017,xu2017,moran2017,meichanetzidisfree}. Local
parafermionic operators have been proposed in hydrid systems combining fractional quantum Hall states
with superconductivity~\cite{lindner2012,cheng2012,clarke2013,vaezi2013,motruck2013,PhysRevB.88.235103,santos2017,alavirad2017,vaezi2017,wu2017,repellin2017}. Their existence, topologically protecting the ground
state (GS) degeneracy, originates from the fractional  edge excitations in quantum Hall states. Improving over Majorana fermions, braiding parafermions
provides access to a richer set of gate operations in quantum computation processing~\cite{hutter2016}.

The existence of local parafermions in truly one-dimensional fermionic systems is still elusive. Symmetry-protected
topological phases of fermions have been classified in one dimension~\cite{Turner2011,fidkowski2011,Bultinck2017} according to
the most common symmetries, and none of them  is found to support edge parafermions.
Despite this classification, parafermions were identified using bosonization in some one-dimensional 
fermionic models~\cite{Klinovaja2014,Klinovaja2014b,Klinovaja2014c,oreg2014,orth2015,pedder2017,vinkler2017}, thus raising two  important questions. The first one is about locality: bosonization
being a non-local transformation, the edge character of these parafermions is not clearly settled. The second
and somewhat related question is whether these parafermions generate topological protection.

In this letter, we delve more deeply into these issues by extending Kitaev's chain~\cite{Kitaev2001} to a lattice fermionic model with $\mathbb{Z}_{N}$ pairing, where $N$ is an even integer. 
Interestingly, the model shares the same bosonization description  with many previous studies where parafermions were identified, and compared to them, it allows for a more microscopic and complete solution through exact results and  numerics.
We find that the GS $N$-fold degeneracy is a mixture of topology and spontaneous symmetry breaking (SSB). We demonstrate
that it is not generated by localized edge parafermions but rather by $N/2$ pairs of Majorana fermions acting in each
symmetry-breaking sector. The system also hosts a pair of non-topological parafermions, called \textit{poor man's parafermions} in Ref.~\onlinecite{kane2015}, that we microscopically demonstrate to be non-local. In fact, it is straightforward to prove that operators, local in terms of fermions, cannot exhibit parafermionic commutation relations (the proof is detailed in~\footnote{See Supplemental Material for more details on bosonization and the exact solution to the GS, the construction of non-local parafermions and edge Majorana modes and the absence of local parafermions in one-dimension fermionic models}), challenging the existence of strong edge parafermions in truly one-dimensional fermionic systems.
Nevertheless, the $N$-fold degeneracy of the GS persists with exponential accuracy for $\mathbb{Z}_{N}$-preserving disorder and the model exhibits the dual of the $8 \pi$ fractional Josephson effect~\cite{zhang2014,orth2015}, thus shading a positive light on the possibility of using non-local and non-topological parafermions in future applications.

\begin{figure}[t]
\includegraphics[width=\columnwidth]{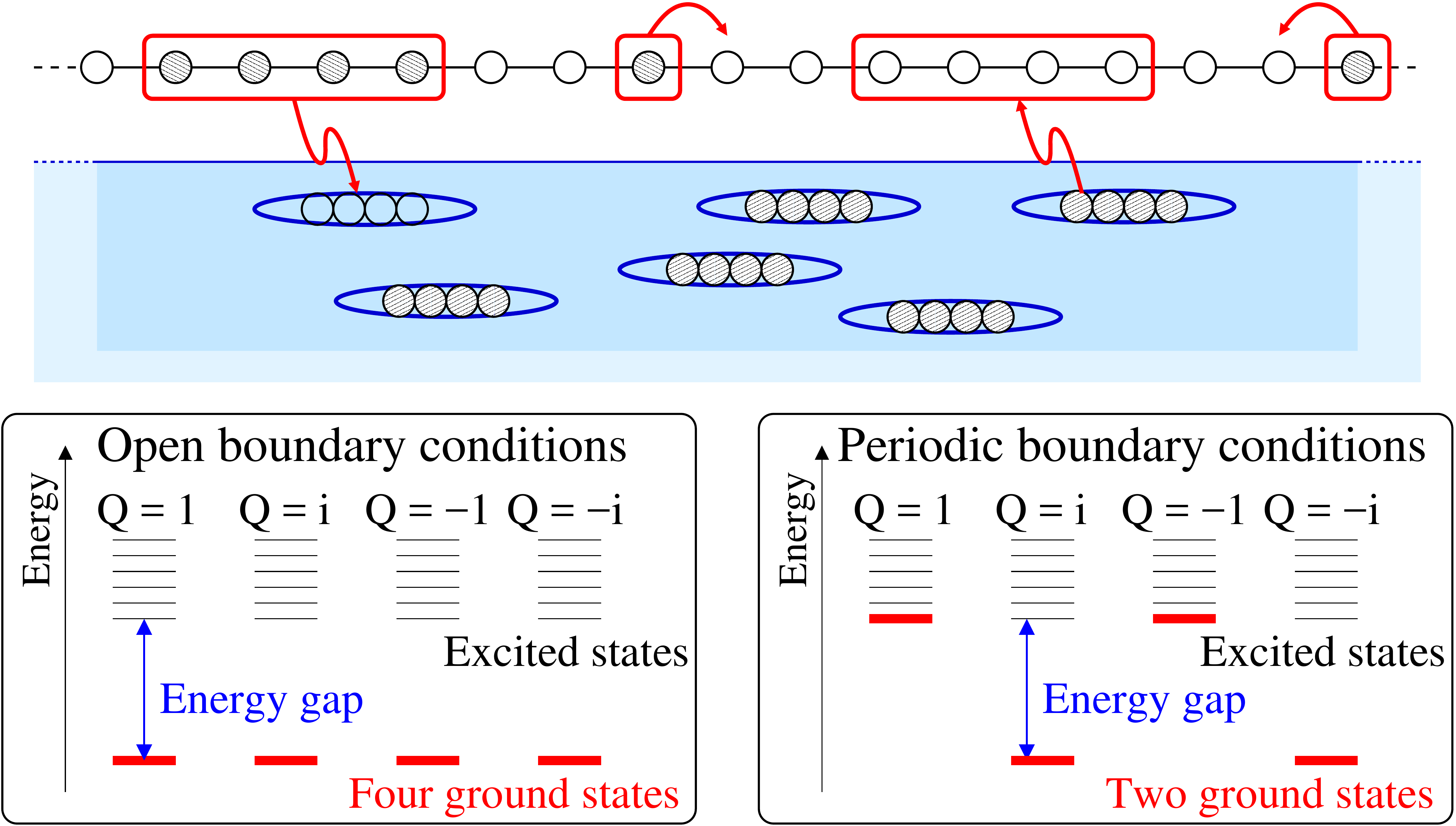}
\caption{Top: Sketch of model~\eqref{hp}. A one-dimensional tight-binding chain of spinless fermions is proximity-coupled to clusters of $N=4$ fermions.
Bottom: Sketch of the spectrum of model~\eqref{hp}. The four-fold degeneracy for open boundary conditions (left) related to the presence of a $\mathbb Z_4$ symmetry, $\hat Q$ (see text), is lifted to a two-fold degeneracy for periodic boundary conditions (right).
}
\label{Fig:sketch}
\end{figure}

\textit{The model ---}
We consider a linear lattice of spinless fermions subject to a multiplet proximity effect involving $N$ fermions 
(the case $N=2$ is simply Kitaev's chain),
\begin{equation}\label{hp}
\hat H_P = - t \sum_j  \hat c_{j+1}^\dagger \hat c_j  + \Delta \sum_j  \left( \prod_{k=0}^{N-1} \hat c_{j+k}^\dagger \right)  + h.c.
\end{equation}
$t$ is the hopping amplitude, $\Delta$ promotes condensation of clusters of $N$ fermions.
 The model breaks the usual $U(1)$ symmetry but possesses a $\mathbb{Z}_{N}$ symmetry, $\hat c_j \to  e^{2 i \pi/N} \hat c_j$, associated to the number of fermions $\hat N_e$ being conserved modulo $N$. The corresponding charge is the generalized parity operator $\hat Q = \omega^{\hat N_e}$, with  $\omega = e^{2 i \pi/N}$, which commutes with the Hamiltonian. 
 
 The low-energy physics is  conveniently described using bosonization~\cite{BosonizationGogolin1998}.  Following standard conventions, we introduce the phase and displacement fields $\hat \theta$ and $\hat \varphi$, where $\partial_x \hat \theta/\pi$ is the momentum conjugated to $\hat \varphi$. The Hamiltonian is written in terms of these bosonic fields $\hat H_P \sim \hat H_0 + \hat H_1$, where
\begin{equation}\label{hamil0}
\hat H_0 =  \frac{v_F}{2 \pi}  \int dx \left[ K (\partial_x \hat \theta)^2 + \frac{1}{K} (\partial_x \hat \varphi)^2 \right] 
\end{equation}
describes the gapless excitations with velocity $v_F$ close to the left and right Fermi points, and $K=1$ in the absence of interactions. The pairing term ($a_0$ is the lattice spacing)
\begin{equation}\label{h1}
\hat H_1 = - \frac{\Delta}{a_0} \int d x \, \cos (N  \hat \theta )
\end{equation}
tends to pin the field $\theta$, increasing its stifness, and thus the Luttinger parameter $K$ in Eq.~\eqref{hamil0}, which makes the system more attractive with increasing $\Delta$. Specifically, the perturbative renormalization group (RG) equations 
\begin{equation}\label{rg}
\frac{d \Delta}{d \ell} = \left(2 - \frac{N^2}{4 K} \right) \Delta, \qquad \frac{d K}{d \ell} = \left( \frac{a_0 \Delta}{v_F} \right)^2 N^2 ,
\end{equation}
with the short length cutoff $a_0 e^\ell$, predict a Kosterlitz-Thouless fixed point~\cite{QP1DGiamarchi2004} at $K = N^2/8$, $\Delta=0$, separating a gapless Luttinger phase from a gapped phase. $\Delta$ flows to strong coupling in the gapped phase and $\hat \theta$ gets locked semiclassically to $2 m \pi/N$, $m$ being an integer. As $\hat \theta$ and $\hat \theta+ 2 \pi$ are identified, there are $N$ distincts ground states following the $\mathbb{Z}_{N}$ symmetry of the model (see Fig.~\ref{Fig:sketch}).

\begin{figure}[t]
 \includegraphics[width=\columnwidth]{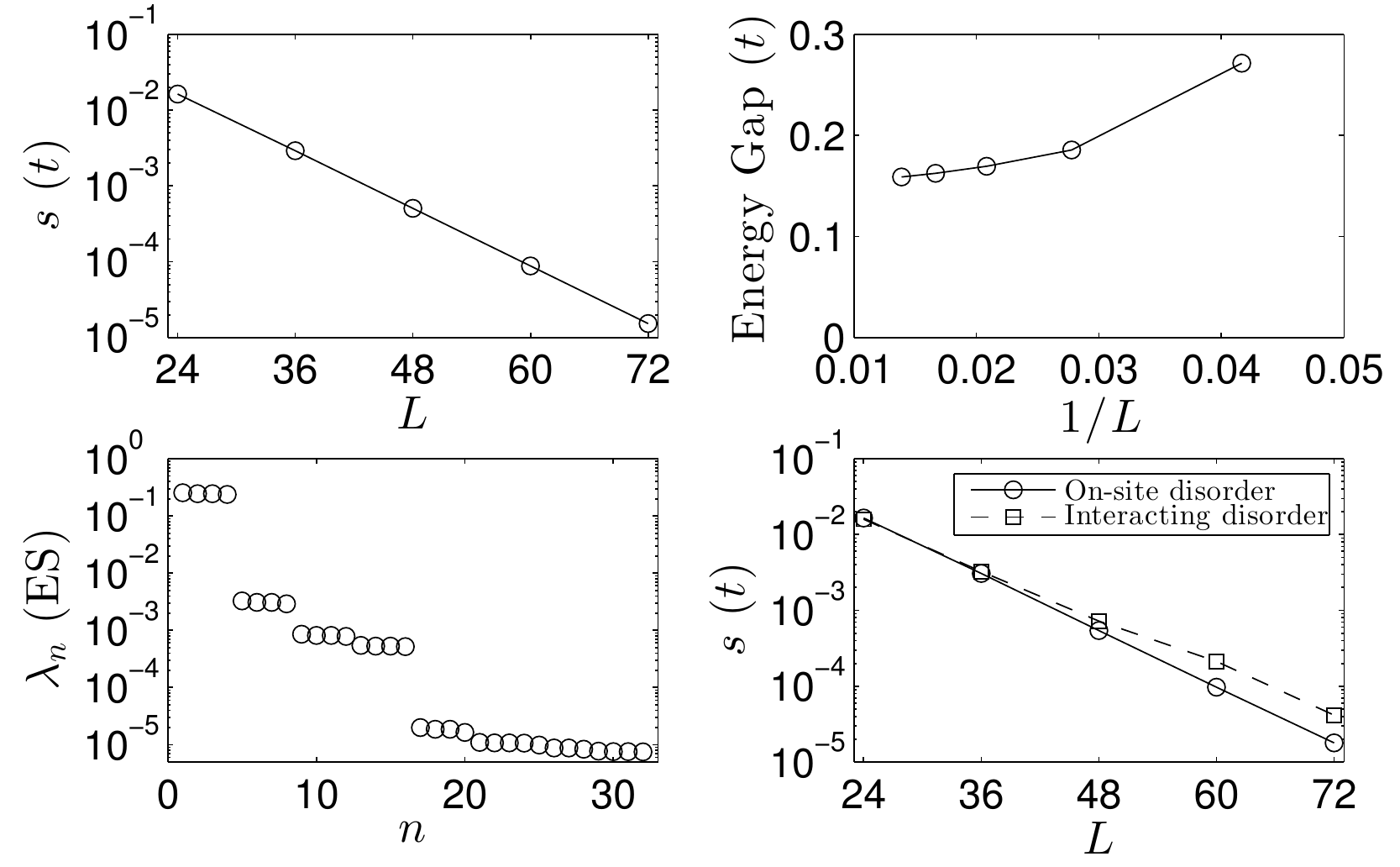}
 \caption{DMRG data of model~\eqref{hp} for $\Delta=t$ with bond link $m = 150$.
 Top-left: standard deviation $s$ of the GS energies for fixed $L$ and different values of $\hat{Q}$.  Top-right: Energy gap between the ground state and the first excited state for $\hat{Q}=1$.
 Bottom-left: Entanglement spectrum for a bipartition of the system of $L=72$ with $\hat{Q} = 1$.
 Bottom-right: standard deviation $s$ of the GS energies for fixed $L$ and different values of $\hat{Q}$ in presence of on-site $\sum_j \delta_j \hat n_j$ and interacting disorder $\sum_j \delta'_j \hat n_j \hat n_{j+1}$. $\delta_j, \delta'_j$ are taken with uniform probability in the range $[-0.1 t, 0.1t]$.}
 \label{Fig:2}
\end{figure}

The properties of the gapped phase can be examined more thoroughly with density-matrix renormalization-group (DMRG) calculations for open chains; in Fig.~\ref{Fig:2} we present data for a representative choice $N=4$, $\Delta = t$.
We find $N$ degenerate GSs protected by a gap (top panels); the entanglement spectrum organizes in degenerate multiplets of $N$ levels  (bottom-left panel).
The exponential energy separation is lifted by disorder terms breaking the  $\mathbb{Z}_{N}$ symmetry and is immune to others. We tested explicitly the cases of an on-site potential $\sum_j \delta_j \hat n_j$ and of a local interaction term $\sum_j \delta'_j \hat n_j \hat n_{j+1}$  (bottom-right panel) preserving $\mathbb{Z}_{N}$. Despite the partial resilience to disorder and the gap induced by Eq.~\eqref{h1} as  in Refs.~\cite{lindner2012,cheng2012,clarke2013}, the GS degeneracy is not entirely topological and the model is not hosting edge parafermions as discussed below.


\textit{Exact solution ---} A better understanding of the gapped phase can be achieved by seeking an exact solution to the model~\cite{iemini2015,lang2015}. This is done by modifying the Hamiltonian $H_P$, {\it i.e.} adding the term
\begin{equation}
  \begin{split}
    \hat H_A = &-t  \sum_j \left[2 \hat n_j \hat n_{j+1} - (\hat n_j + \hat n_{j+1}) \right]  \\
    &+ \Delta \sum_j  \left[ \prod_{k=0}^{N-1} \hat n_{j+k} + \prod_{k=0}^{N-1} \left(1 - \hat n_{j+k} \right) \right],
  \end{split}
\end{equation}
such that its ground states can be explicitly determined. One first verifies that the Hamiltonian $H_P + H_A$ is a non-negative operator. We then observe that for an open chain of size $L$, the following states are zero-energy eigenstates
  \begin{equation}\label{eigenstates}
    \ket{\psi_{p}} = \sqrt{\frac{N}{2^L}} \sum_{j_1 < \ldots <j_{N_e} \atop N_e = p[N]} \ket{\{ j_{N_e} \}}.
\end{equation}
They  correspond to the coherent sum~\cite{greiter2014,Katsura2015} of all Fock states $\ket{\{ j_{N_e} \}} = \hat c_{j_1}^\dagger \hat c_{j_2}^\dagger \ldots \hat c_{j_{N_e}}^\dagger \ket{0}$  such that the number of fermions $N_e$ is equal to  $p$  modulo $N$. 
Since $\hat H_P + \hat H_A$ is non-negative, the $N$ states $\ket{\psi_{p}}$, with $0 \le p \le N-1$, belong to the GS manifold (we have verified numerically that they exhaust it). As they are eigenvalues of the $\mathbb{Z}_{N}$ parity operator,  $\hat Q  \ket{\psi_{p}} = \omega^p \ket{\psi_{p}}$, each parity sector has a unique zero-energy GS.

\begin{figure}[t]
 \includegraphics[width=\columnwidth]{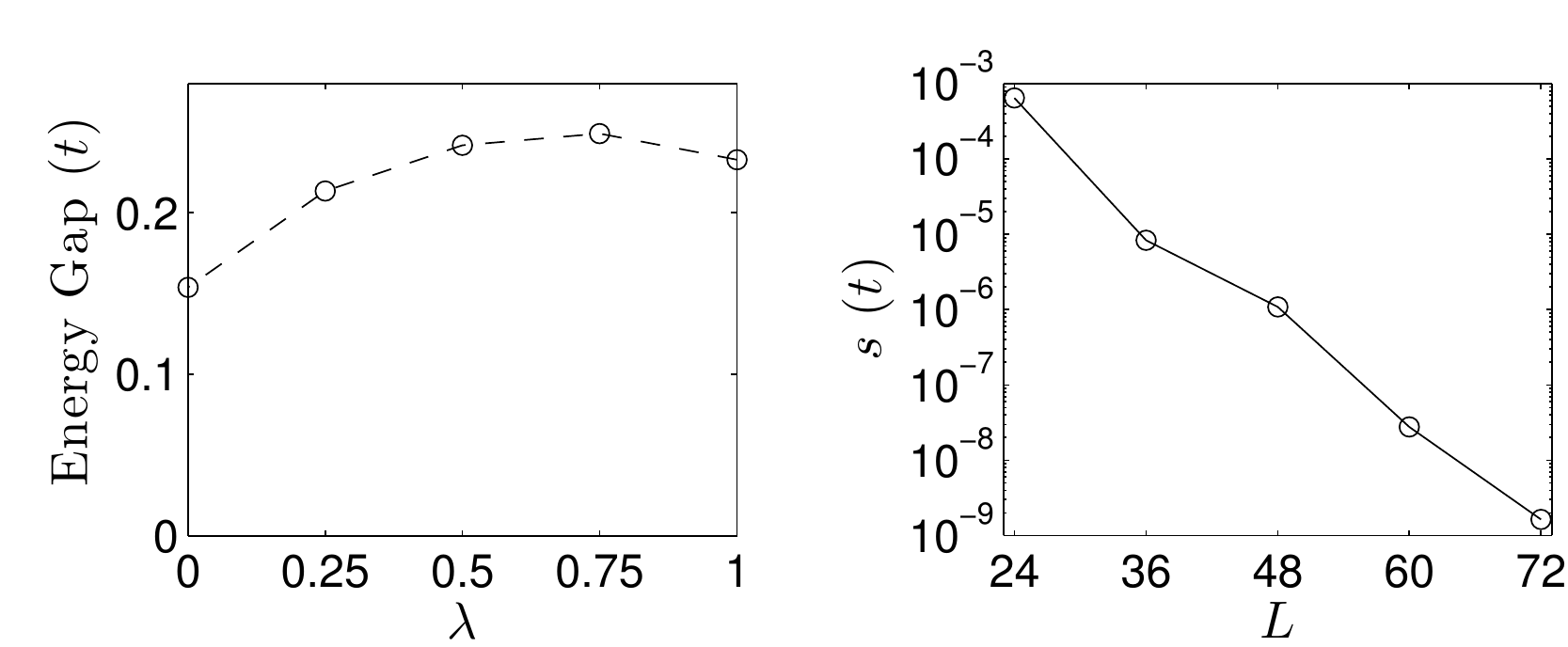}
\caption{DMRG data of the model $\hat H_P + \lambda \hat H_A$ ($m=150$). Left: Gap of the model as a function of $\lambda$ extracted with a finite-size scaling for $L$ between $24$ and $72$. Right: standard deviation of the GS energies at $\lambda = 0.5$ for fixed $L$ and different values of $Q$. A small on-site disorder $\sum_j \delta_j \hat n_j$ is introduced with $\delta_j \in [-0.1 t, 0.1 t]$.}
\label{Fig:3}
\end{figure}

Before studying the properties of these states, we need
to ensure that they describe the same gapped phase of $\hat H_P$ (or, equivalently, of $\hat H_0 + \hat H_1$). 
We use a concrete numerical strategy by building an adiabatic path relating the two models $\hat H_P$ and $\hat H_P + \lambda \hat H_A$. 
In Fig.~\ref{Fig:3}  we monitor the gap of the model along the path $\lambda \in [0,1]$ and test the GS degeneracy and resilience to disorder for $\lambda = 0.5$. This demonstrates by continuity the uniqueness of the gapped phase and thus the relevance of the analytical states $\ket{\psi_{p}}$ to describe it.

\textit{Topological properties of the exact model ---}
Topological phases typically host boundary zero modes in open geometries which induce circular permutations among the degenerate GS. The presence of these modes on the edges reflects the insensitivity of the bulk to perturbations. Closing the chain lifts entirely the GS degeneracy by coupling the edge states~\cite{Alexandradinata2016}. This situation contrasts with a non-topological phase with SSB, in which case closing the chain does not change the degeneracy. 

In order to test the topological properties of our model, we consider the Hamiltonian $\hat H_P + \hat H_A$ on a  chain with periodic boundary conditions. The states $\ket{\psi_{p}}$ with odd $p$ remain ground states with zero energy; those with even $p$ frustrate at least one link between lattice sites and thus have energy larger than zero. 
The $N$-fold GS degeneracy is thus reduced to $N/2$-fold  when closing the chain (see Fig.~\ref{Fig:sketch}).
This partial reduction unambiguously indicates that the GS degeneracy is only partially topological~\cite{bondesan2013,motruck2013}.  

As we show below, the partial lift of degeneracy can be accounted for by the presence of edge Majorana fermions. The residual $N/2$ degeneracy, instead, comes from SSB  induced by the operator
\begin{equation}\label{o2}
\hat{\cal O}_2^{(j)} = 2 e^{-i \varphi} \, \hat c_j^\dagger \hat c_{j+1}^\dagger + h.c.,
\end{equation}
where $\varphi$ is an arbitrary phase. 
The peculiar mix of topology and SSB clearly appears once the matrix elements $\bra{\psi_{p+1}} \hat c_j^\dagger \ket{\psi_p}$, exponentially small in the bulk, and $\bra{\psi_{p+2}} \hat c_j^\dagger\hat c_{j+1}^\dagger \ket{\psi_p}=1/4$ are analytically computed.
In parafermionic chains such mixing of topology and SSB is resulting from pair condensation of parafermions~\cite{bondesan2013,motruck2013}.

Restricting operators~\eqref{o2} to the GS manifold spanned by the states~\eqref{eigenstates}, they are diagonalized by the states
\begin{equation}\label{gsn}
  \ket{n,e} =  \sum_{q=0}^{\frac N2-1} \frac{\omega^{2qn}}{\sqrt{N/2}} \ket{\psi_{2q}},
  \;\;
  \ket{n,o} = \sum_{q=0}^{\frac N2-1} \frac{\omega^{ (2q+1)n}}{\sqrt{N/2}} \ket{\psi_{2q+1}};
\end{equation}
$\sigma=e,o$ denoting the fermionic parity, which spontaneously break the $\mathbb{Z}_{N}$ symmetry down to a topological $\mathbb Z_2$ symmetry. This is shown by the eigenvalues ${\cal O}_2^{(j)}  \ket{n,\sigma} =  \cos \left( \varphi - 4 \pi n/N \right)  \ket{n,\sigma}$ independent on $\sigma$.

A further insight into the fact that the pairs $\ket{n,e}$ and $\ket{n,o}$ are protected by topology is given by the observation that
they are also the exact ground states of the Kitaev models ($n=1,\ldots,N/2$)
\begin{equation}
\hat H_{K,n} = -t_0  \sum_j  \left( \hat c_j^\dagger \hat c_{j+1} +  \omega^{2 n} \,  \hat c_j^\dagger  \hat c_{j+1}^\dagger + h.c. \right)
\end{equation}
related to each other by the circular gauge transform $\hat c_j \to \omega \, \hat c_j$. A direct consequence is that each of the following pairs of edge Majorana operators
\begin{equation}\label{MajoES}
\hat \alpha_{1,n} = \omega^{-n} \hat c_1 + \omega^{n} \hat c_1^\dagger \qquad \hat \alpha_{2,n} = i ( \omega^{-n} \hat c_L - \omega^{n} \hat c_L^\dagger ),
\end{equation}
generates the algebra of Pauli matrices when restricted to the states $\ket{n,e}$ and $\ket{n,o}$.  

To summarize the above discussion, the GS manifold splits into $N/2$ symmetry-breaking sectors. In each sector, a pair of Majorana fermions enforces a topological degeneracy which is lifted when closing the chain while a residual $N/2$ degeneracy is preserved.

\textit{Topological properties through bosonization ---}
To confirm the GS structure beyond the fine-tuned Hamiltonian $\hat H_P + \hat H_A$, we return to the bosonized form $\hat H_0+ \hat H_1$ with a chain of finite length $L$ and periodic boundary conditions. We determine the size of the GS manifold through
  the mode expansion
\begin{equation}\label{modeexpansion}
\begin{split}
\hat \theta (x) &=  \hat \theta_0 + \frac{\pi x}{L} \hat J  + \hat \theta_{k \ne 0} (x) \\[1mm]
\hat \varphi (x) &=  \hat \varphi_0 + \frac{\pi x}{L} \hat N_e + \hat \varphi_{k \ne 0}  (x)
\end{split}
\end{equation}
that decomposes the bosonic fields into zero modes and finite momentum excitations. The phase variables $\hat \varphi_0$ and $\hat \theta_0$ are conjugate variables to the (integer-valued) current $\hat J$ and particle number $\hat N_e$~\footnote{More precisely, the number of fermions is $N_0 + \hat N_e$, where $N_0$ is a reference value.},
\begin{equation}\label{zero}
[\hat \varphi_0, \hat J] = - i, \qquad \qquad [\hat \theta_0, \hat N_e] = - i,
\end{equation}
with the gluing (parity) condition~\cite{Haldane1981} $(-1)^{\hat N_e} =  (-1)^{\hat J}$. 
Substituting the expansion~\eqref{modeexpansion}  into the Hamiltonian $\hat H_0+ \hat H_1$, we get that in the limit of very large pairing $\Delta$ the current must vanish $\hat J = 0$ for $\hat \theta$ to be spatially uniform. 
The gluing condition then enforces that $\hat N_e$ is an even integer restricting the GS manifold to $N/2$ states in agreement with the exact-solution analysis.

Next, we consider the gapped phase in the open geometry with a vanishing particle current on both ends, $\partial_x \hat \theta (x=0,L)=0$. In this case, the mode expansion is simplified and $\hat J$ and $\hat \varphi_0$ are removed from Eq.~\eqref{modeexpansion}. The minima of the cosine in $\hat H_1$  are the eigenstates $\ket{\theta_m}$ of $\hat \theta_0$ with eigenvalues $2 m \pi/N$ ($m=0,1 \ldots,N-1$). These states are not eigenstates of the generalized parity operator $\hat Q$,  $\hat Q \ket{\theta_m} = \ket{\theta_{m+1}}$ from Eq.~\eqref{zero}. They are coupled at finite size $L$ by tunneling of instantons between the different minima of the cosine potential in Eq.~\eqref{h1}. Such tunnelings transfer a charge $\Delta \theta/2 \pi = 1/N$ between the two chiral left-moving and right-moving channels~\cite{qi2008}.
The GS eigenstates of $\hat Q$ are given by the linear combinations~\cite{seidel2005} $\ket{\Psi_p} = \sum_{m=0}^{N-1} \omega^{-m p}  \ket{\theta_m}$, in agreement with~\eqref{eigenstates} and~\eqref{gsn}~\footnote{a tight analogy requires the identification  \unexpanded{$| n,e  \rangle  + | n,o  \rangle  \to | \theta_n \rangle$} and  \unexpanded{$| n,e  \rangle  - | n,o  \rangle  \to | \theta_{n+N/2} \rangle$}}.
Their energy splitting is exponentially small with $L$ as a result of instanton tunneling~\cite{Fisher2011}.



Having set the bosonization framework, we reconsider the resilience of the gapped topological phase to disorder made of local $\mathbb{Z}_{N}$-preserving operators, as displayed in Fig.~\ref{Fig:2}. It amounts to test whether the different GS can be distinguished by such operators. $\mathbb{Z}_{N}$-preserving operators have a bosonized form involving either the field $\hat \phi$, derivatives of $\hat \theta$ or  $e^{i N \theta}$. They cannot measure the zero mode $\hat \theta_0$ and thus discriminate the different GS~\cite{cheng2011}. By contrast, the bosonized form of the operators~\eqref{o2}
\begin{equation}
\hat{\cal O}_2^{(j)}  \sim \cos (\varphi- 2 \hat \theta), \quad \hat{\cal O}_2^{(j)}  \ket{\theta_m} =  \cos \left( \varphi - 4 \pi m/N \right) \ket{\theta_m}
\end{equation}
clearly discriminate the GS manifold into $N/2$ sectors in full agreement with the exact solution. The $N$-fold degeneracy of the model~\eqref{hp} is therefore symmetry-protected~\cite{zhang2014} by number conservation modulo $N$. Even when pairing terms lift the GS degeneracy, a twofold degeneracy protected by edge Majorana fermions is nonetheless guaranteed.


\textit{Parafermions ---}
At this stage, it might appear surprising to discover that the gapped phase hosts parafermionic zero modes
\begin{equation}\label{paraf}
\hat \gamma_1 = e^{i \hat  \theta_0}, \qquad \hat \gamma_2 = e^{i \left( \hat \theta_0 + \frac{2 \pi \hat N_e}{N}\right)}.
\end{equation}
$\hat \theta_0$ is a phase variable with the identification $\hat \theta_0 \sim \hat \theta_0 + 2 \pi$ so that
$e^{i \hat  \theta_0}$ is a legitimate operator. $\hat \theta_0$ is pinned at low energy in the minima of Eq.~\eqref{h1} and can be written as  $\hat \theta_0 = 2 \pi \hat n_\theta/N$. $\hat n_\theta$ is an integer-valued operator which does not commute with $\hat N_e$, as Eq.~\eqref{zero} indicates. As a result, $\hat \gamma_1$ and $\hat \gamma_2$ satisfy the properties expected for parafermions, namely $\hat \gamma_1^N = \hat \gamma_2^N = 1$ and $\hat \gamma_1  \hat \gamma_2 = \omega \hat \gamma_2 \hat \gamma_1$. They commute with the Hamiltonian $\hat H_0 + \hat H_1$ and therefore define zero-energy modes permuting the whole GS manifold with $N$ states. Corresponding parafermionic operators are also constructed from the exact solution~\cite{Note1}.   Both constructions highlight the fact that these operators are non-local, {\it i.e.} not edge properties of the chain. For this reason, their existence does not contradict the partially topological nature of the gapped phase.
As such, the degeneracy that they describe is not topological: they correspond to the \textit{poor man's parafermions} discussed in Ref.~\cite{kane2015}.

These non-local parafermions are beyond a nice mathematical construction: the system exhibits an anomaly very similar to the $8 \pi$ Josephson effect predicted for quantum spin Hall edge states~\cite{zhang2014,orth2015}, as the two models are dual to each other in bosonization. Let us consider the open chain for $N=4$ connected on each side to a 1D band (Peierls) insulator. The insulators are described in bosonization by the gapping term $\cos(2 \hat \phi - \varphi_{1/2})$  in addition to the quadratic part~\eqref{hamil0}, preserving the $\mathbb{Z}_{4}$ symmetry. $\varphi_1$ ($\varphi_2=0$) is the phase acquired in the left (right) insulator upon backscattering a right-moving to a left-moving single fermion. 
  As $\theta_m \to \theta_{m+1}$ tunneling events in the multiplet pairing region involve fractional fermion backscattering of charge $1/4$, their amplitudes are dressed by $e^{i \varphi_1/4}$. Therefore, although the Hamiltonian is invariant upon shifting $\varphi_1$ by $2 \pi$, the spectral flow of the splitted GS manifold is $8 \pi$ periodic in  $\varphi_1$ and level crossings are protected by $\mathbb{Z}_{4}$ symmetry, {\it i.e.} fermion number conservation modulo $4$. The case $N=2$, in the context of the magneto-Josephson effet, was solved exactly~\cite{meng2012,liang2013,pientka2013}.

Following Ref.~\cite{clarke2013}, it is also possible to introduce \textit{local} boundary zero-mode operators, {\it i.e.} commuting with the Hamiltonian, sitting on the edges of the open chain. They take the form
\begin{equation}\label{majoranas}
\hat \beta_1 \sim e^{i \hat  \theta_0}, \qquad \hat \beta_2 \sim e^{i \left( \hat  \theta_0 +\pi \hat N_e \right)}.
\end{equation}
Although $\hat \beta_1^N = \hat \beta_2^N =1$,  the fermionic statistics $\hat \beta_1 \hat \beta_2 = - \hat \beta_2 \hat \beta_1$ implies that these operators permute the pair of states $\ket{n,e}$ and $\ket{n,o}$, see Eq.~\eqref{gsn}, but do not couple the SSB sectors with different $n$. In each SSB sector, $\hat \beta_1$ and $\hat \beta_2$ are identified as the Majorana fermions $\hat \alpha_{1,n}$ and $\hat \alpha_{2,n}$ (Eq.~\eqref{MajoES}) introduced with the exact solution. The comparison between the non-local parafermions~\eqref{paraf} and the local edge (Majorana) modes~\eqref{majoranas} reveals that boundary operators, such as $\hat \beta_2$, can distinguish~\cite{Turner2011} at most the fermionic parity $e^{i \pi \hat N_e}$. This contrasts with operators sensitive to the  $\hat Q$-parity $e^{i \frac{ 2 \pi \hat N_e }{N}}$, such as $\hat \gamma_2$, but necessarily delocalized on the whole chain.

\textit{Conclusions ---}
 We  discussed a microscopic one-dimensional fermionic model characterized by even multiplet pairing, in many aspects similar to models put forward as candidates for realizing parafermions in one-dimensional fermionic systems.
 We explicitly constructed the parafermions of our model and shown that they are neither local nor topological. The peculiar discussed phenomenology is instead a mixture of topological Majorana physics and SSB. We believe that these results extend beyond our model. However, the eccentricity of the specific SSB that we found might still leave non-topological parafermions as a viable route for technological applications.

  \textit{Note added.}
 After the upload of this article on the arXiv website, two other independent preprints appeared that, discussing different models and using related techniques, analyzed the phenomenology that is the object of this paper, namely that of topological Majorana physics endowed with spontaneous symmetry breaking~\cite{Chew2018, Calzona2018}.

\textit{Acknowledgements ---}
We thank A.~Bernevig, J.~Budich, M.~Burrello, R.~Fazio, D.~Loss, N.~Regnault, E.~Sela, A.~Stern and P.~Zoller for enlightening discussions.
L.M was supported by LabEX ENS-ICFP: 
ANR-10-LABX-0010/ANR-10-IDEX-0001-02 PSL*. C.~M.~acknowledges support from Idex PSL Research University (ANR-10-IDEX-0001-02 PSL).

\bibliography{pairing}

\clearpage
\newpage

\clearpage
\setcounter{equation}{0}%
\setcounter{figure}{0}%
\setcounter{table}{0}%
\renewcommand{\thetable}{S\arabic{table}}
\renewcommand{\theequation}{S\arabic{equation}}
\renewcommand{\thefigure}{S\arabic{figure}}

\onecolumngrid

\begin{center}
  {\Large Supplemental Material for: \\ \mytitle}

\vspace*{0.5cm}
 Leonardo Mazza, Fernando Iemini, Marcello Dalmonte and Christophe Mora
\vspace{0.25cm}
\end{center}



The different analytical techniques mentionned in the main text are presented with more details in this Supplementary information.
Three topics are covered: the formulation of an exact solution for the ground state at a fine-tuned point, the low-energy bosonisation
approach and the generic demonstration that parafermionic statistics cannot be obtained from spatially separated operators
in one-dimensional fermionic systems.

\section{Exact solution for the ground states}\label{sec:exact}

\subsection{Number-conserving Hamiltonian}

The special Hamiltonian advertised in the main text, $H_P + H_A$ can be decomposed into two contributions,  $H = H_P + H_A = H_{Hei} + H_{pairing}$. With $\hat c_j$ the operator annihilating a spinless fermion at site $j$ and $\hat n_j = \hat c_j^\dagger \hat c_j$ the corresponding on-site density, the first term in $H$,
\begin{equation}
H_{Hei}  = -t  \sum_{j=1}^{L-1}  \left[ \left( \hat c_j^\dagger \hat c_{j+1} + {\rm H.c.} \right) - (\hat n_j + \hat n_{j+1}
) +  2 \hat n_j \hat n_{j+1} \right],
\end{equation}
maps onto the ferromagnetic Heisenberg model upon a Jordan-Wigner transform. The number of fermions $\hat N_e$ is conserved with this Hamiltonian. Also by construction $H_{Hei}  \ge 0$ and a zero-energy ground state is found in each sector with fixed $\hat N_e = N_e$. The ground state wavefunctions take the form
 \begin{equation}\label{states}
 \ket{\psi_{N_e}} = \frac{1}{\sqrt{\mathcal N_{N_e}}}
 \sum_{ 1 \leq j_1 < j_2 < \ldots <j_{N_e} \leq L} \ket{\{ j_{N_e} \}},
 \qquad \mathcal N_{N_e} = \binom{L}{N_e}
\end{equation}
corresponding to a coherent sum of Fock states 
$$\ket{\{ j_{N_e} \}} = \hat c_{j_1}^\dagger \hat c_{j_2}^\dagger \ldots \hat c_{j_{N_e}}^\dagger \ket{0},$$ $\ket{0}$ denoting the vacuum state. Already at this level, it is possible to distinguish different boundary conditions. For open boundary conditions, the states \eqref{states} are genuine zero-energy ground states of the positive Hamiltonian $H_{Hei}$. However, for periodic or antiperiodic boundary conditions $\hat c_{j+L} = \pm  \hat c_{j}$, only the states \eqref{states} with $N_e$ odd/even are ground states of $H_{Hei}$.

\subsection{Adding multiplet pairing}

We now add the multiplet pairing contribution to the Hamiltonian corresponding to the proximity hopping of $N$ fermions,
\begin{equation}
  H_{pairing} = \Delta \sum_{j=1}^{L-N+1} \left[ \left( \prod_{k=0}^{N-1} \hat c_{j+k}^\dagger  + h.c. \right) + \prod_{k=0}^{N-1} \hat n_{j+k} + \prod_{k=0}^{N-1} \left(1 - \hat n_{j+k} \right) \right]
\end{equation}
$H_{pairing}$ is a  positive operator. Consider $A_j = B_j + B_j^2$ with $B_j = \prod_{k=0}^{N-1} \hat c_{j+k}^\dagger  + h.c.$. Noticing that $B_j^3 = B_j$, one easily obtains  $A_j^2 = 2 A_j$  which implies that the eigenvalues of $A_j$ are either $0$ and $2$. $H_{pairing} = \Delta \sum_j A_j$ is therefore a sum of positive operators and itself a positive operator. The total Hamiltonian $H = H_{Hei} + H_{pairing}$ does not conserve the number of fermions,  $[\hat H,\hat N_e] \ne 0$, but only the $Z_N$ parity, $[\hat H,\hat Q] = 0$, where $\hat Q = \omega^{\hat N_e}$ and $\omega = e^{2 i \pi/N}$.

The Hamiltonian has a complicated form but, most importantly, it is local in terms of the fermionic operators. It has analytical eigenstates at zero energy. Among the states~\eqref{states}, there are $N$ linear combinations which minimize both $H_{Hei}$ and $H_{pairing}$, namely
\begin{equation}\label{eigenstates2}
    \ket{\psi_{p}} = \sqrt{\frac{N}{2^L}} \sum_{j_1 < \ldots <j_{N_e} \atop N_e = p[N]} \ket{\{ j_{N_e} \}}, \qquad \qquad p=0,\ldots, N-1
\end{equation}
with $H_{Hei} \ket{\psi_{p}} = H_{pairing} \ket{\psi_{p}} = 0$. These states form the ground state manifold of $H$. Following the above analysis, for periodic or antiperiodic boundary conditions, only the states with $p$ odd/even are ground states of $H$.

\subsection{Symmetry breaking operator and degeneracy lifting}

Part of the degeneracy is the result of symmetry breaking just as in the quantum Ising model. The corresponding local operator involves the hopping of a pair of fermions (measuring p-wave pairing),
\begin{equation}\label{o2-2}
{\cal O}_2^{(j)} = e^{i \varphi} \, \hat c_j^\dagger \hat c_{j+1}^\dagger + h.c.,
\end{equation}
decoupling the odd and even particle number sectors. $\varphi$ here is an arbitrary phase. Restricting ${\cal O}_2^{(j)}$ to the ground state manifold with $p$ odd, it takes the form, for instance when $N=6$,
\begin{equation}
{\cal O}_{2}^{(j)} = \frac{1}{4} \begin{pmatrix} 0 & e^{-i \varphi} & e^{i \varphi} \\ e^{i \varphi} & 0 & e^{-i \varphi} \\ e^{-i \varphi} & e^{i \varphi} & 0
\end{pmatrix},
\end{equation}
and the same reduced form for $p$ even. Hence, it is easily diagonalized as shown in the main text. Adding a coupling to the local operator ${\cal O}_2^{(j)}$ to the Hamiltonian would lift the degeneracy between the different symmetry-breaking sectors. Each of these sectors still preserves a two-fold degeneracy, {\it i.e.} a ground state manifold of dimension two, associated to the even/odd parity and to the presence of Majorana edge modes (see the corresponding discussion in the main text). This residual degeneracy is topologically protected.

\subsection{Non-local Parafermions}\label{sec:paraf}

With the knowledge of the exact wavefunctions, it is possible to define operators which behave as parafermions when restricted to the ground state manifold. To construct these operators, we first introduce the notation
\begin{equation}
\tilde c_j  = \hat c_j e^{i \pi \sum_{k<j} \hat n_k},
\end{equation}
corresponding to an annihilation operator dressed by the standard Jordan-Wigner string~\footnote{In the spin language, $\tilde c_j$  corresponds to the operator $S_j^-$ flipping the local spin from up to down.}. The string implies that $\tilde c_j$ commutes with all fermionic operators located on its left,  {\it i.e.} $[\tilde c_j,\hat c_k] = [\tilde c_j,\hat c_k^\dagger] = 0$ for $k<j$.

Using this commutation property, we introduce the set of all operators removing one fermion (modulor $N$)  from the chain with the ordered products
\begin{equation}
 {\cal K}_{\vec{\ell},\vec{k}} = \tilde c_{\ell_1}^\dagger \ldots \tilde c_{\ell_{N_+}}^\dagger  \tilde c_{k_1} \ldots \tilde c_{k_{N_-}} \qquad \ell_1 < \ldots < \ell_{N_+} \qquad k_1 < \ldots < k_{N_-} 
\end{equation}
and $N_+ - N_- = -1 [N]$. For any pair of Fock states $\ket{\{ j_{N_e} \}}$ and $\ket{\{ j_{N_e'} \}}$ with $N_e' = N_e-1 [N]$, there exists a unique operator ${\cal K}_{\vec{\ell},\vec{k}} $ such that 
$\ket{\{ j_{N_e'} \}} = {\cal K}_{\vec{\ell},\vec{k}} \ket{\{ j_{N_e} \}}$. Hence, summing over the whole set of ${\cal K}_{\vec{\ell},\vec{k}} $ operators, we define a pair of non-local operators 
\begin{equation}\label{comm}
  \hat \gamma_1 =  \frac{N}{2^L} \sum_{\vec{\ell},\vec{k}} {\cal K}_{\vec{\ell},\vec{k}}, \qquad \qquad \hat \gamma_2 = \hat \gamma_1 e^{\frac{2 i \pi  (\hat N_e-1/2)}{N}}, \qquad \qquad \hat \gamma_1 \hat \gamma_2 =  \omega \, \hat \gamma_2 \hat \gamma_1.
\end{equation}
In terms of the fermionic degrees of freedom, the domain of existence of $\hat \gamma_1$ and $\hat \gamma_2$ coincides with the full chain. Acting on the ground states, one finds
\begin{equation}
  \hat \gamma_1  \ket{\psi_{p}} = \ket{\psi_{p-1}}  \qquad \qquad  \hat \gamma_2  \ket{\psi_{p}} = \omega^{p-1/2} \ket{\psi_{p-1}}
\end{equation}
This implies $\hat \gamma_1^N =  \hat \gamma_2^N=1$ within the ground state manifold which, together with the commutation relation~\eqref{comm}, shows that $\hat \gamma_1$ and $\hat \gamma_2$ behave as parafermions encoding the ground state degeneracy.

\section{Bosonization}

We discuss below the details of the bosonization method applied to the generalized pairing model.

\subsection{Notations}

We assume that interactions and pairing are weak enough and consider only the kinetic energy. We  take the continuum limit of the lattice model and project the fermion field operator around the two Fermi points
\begin{equation}
\hat c_j \simeq \sqrt{a_0} \left( \hat \psi_R (x) e^{i k_F x} + \hat \psi_L (x) e^{-i k_F x} \right),
\end{equation}
at $x = j a_0$. We bosonize the two fields 
\begin{equation}
\hat \psi_{R/L} (x) = \frac{1}{\sqrt{2 \pi a_0}} e^{i [\hat \theta(x) \pm \hat \varphi (x)]},
\end{equation}
 $a_0$ is the short-distance cutoff or lattice spacing (Klein factors are discarded here for simplicity as they play no role in the forthcoming discussion). The two bosonic fields $\hat \theta$ and $\hat \varphi$ satisfy the canonical commutation relation
\begin{equation}
[ \partial_x \hat \varphi (x) , \hat \theta(x') ] = i \pi \delta (x-x').
\end{equation}
We now add interactions and pairing, and analyze the model by the perturbative renormalization group method. If we suppose that the fermion interactions are overall attractive, the Luttinger parameter takes a value $K >1$. Cosine terms involving the field $\hat \varphi$ flow to zero under RG and are thus irrelevant. Rescaling the coupling $\Delta$ to absorb numerical coefficients of order $1$, we obtain
\begin{equation}\label{bos}
H_{bos} = \frac{v_F}{2 \pi}  \int dx \left[ K (\partial_x \hat \theta)^2 + \frac{1}{K} (\partial_x \hat \varphi)^2 \right] - \frac{\Delta}{a_0} \int d x \, \cos (N  \hat \theta ),
\end{equation}
written as $H_0+H_1$ in the main text. The RG equations are given as Eq.~\eqref{rg} in the main text.

\subsection{Boundary conditions}

We first consider periodic boundary conditions and expand the bosonic fields over the mode expansions, written as Eqs~\eqref{modeexpansion} in the main text,
\begin{equation}\label{mode-exp}
\begin{split}
\hat \varphi (x)  & = \hat \varphi_0 + \frac{\pi x}{L} \hat N_e + \frac{i \pi}{L} \sum_{k \ne 0}  \left( \frac{L |k|}{2 \pi} \right)^{1/2} \frac{1}{k} e^{- a_0 |k|/2 - i k x} ( \hat a_k^\dagger + \hat a_{-k} ), \\
\hat \theta (x)  & = \hat \theta_0 + \frac{\pi x}{L} \hat J + \frac{i \pi}{L} \sum_{k \ne 0}  \left( \frac{L |k|}{2 \pi} \right)^{1/2} \frac{1}{|k|} e^{- a_0 |k|/2 - i k x} ( \hat a_k^\dagger - \hat a_{-k} ), 
\end{split}
\end{equation}
where $k$ are multiples of $2 \pi /L$, and the phase variables $\hat \varphi_0$ and $\hat \theta_0$ are conjugate variables to the current $\hat J$ and particle number $\hat N_e$,
\begin{equation}\label{zero2}
[\hat \varphi_0, \hat J] = - i, \qquad \qquad [\hat \theta_0, \hat N_e] = - i.
\end{equation}
In addition, the periodicities of the fields $\hat \psi_{R/L}$ imply the selection rule $(-1)^{\hat N_e} = (-1)^{\hat J}$. When relevant, the pinning of the field $\hat \theta$ in the minima of the cosine in Eq.~\eqref{bos} imposes that $\hat \theta$ is spatially homogeneous, thus discarding a non-zero value for $\hat J$ in Eq.~\eqref{mode-exp}. From the gluing condition $(-1)^{\hat N_e} = (-1)^{\hat J}$, we obtain that the ground state must have an even $\hat N_e$.

If instead we consider an open chain with a vanishing current at both ends,  $\partial_x \hat \theta (x=0,L)=0$, the mode expansion takes a simplier form
\begin{equation}\label{mode-exp2}
\begin{split}
\hat \varphi (x)  & = \hat \varphi_0 + \frac{\pi x}{L} \hat N_e + i \sum_{k > 0}  \left( \frac{\pi}{L k}  \right)^{1/2} \sin (k x) \, ( \hat a_k^\dagger + \hat a_{k} ), \\
\hat \theta (x)  & = \hat \theta_0  + i \sum_{k > 0}  \left( \frac{\pi}{L k}  \right)^{1/2} \cos( k x) \, ( \hat a_k^\dagger - \hat a_{k} ), 
\end{split}
\end{equation}
$k$ are multiples of $\pi /L$ and the canonical commutation relations $[\hat a_k,\hat a_{k'}^\dagger] = \delta_{k,k'}$, $ [\hat \theta_0, \hat N_e] = - i$ apply.
Here all values of the particle number $\hat N_e$ are admissible and the ground state manifold has dimension $N$ in contrast to the closed chain.

\subsection{Parafermionic zero modes}

Substituting the expansion~\eqref{mode-exp2} into the Hamiltonian~\eqref{bos} (or $H_0+H_1$ in the main text), one finds that the operator $\hat \theta_0$ is pinned in the gapped phase to the minima of the cosine. This is taking into account by writting  $\hat \theta_0 = 2 \pi \hat n_\theta/N$ where $\hat n_\theta$ is an integer-valued operator. Assuming at low energy that the modes $\hat a_k$ have a perturbative effect, we expand the Hamiltonian to isolate the part which involves the zero modes
\begin{equation}\label{zeromodeha}
  H_{bos} = \frac{\pi v_F}{2 K L} \hat N_e^2 - \frac{\Delta}{a_0} L \cos ( N \hat \theta_0 ) + \ldots
\end{equation}
where the cosine can also be written $\cos ( N \hat \theta_0 ) = \cos(2 \pi \hat n_\theta)$. The $Z_N$ parity operator $\hat Q = \omega^{\hat N_e}$ shifts the value of $\hat n_\theta$ by one units, $\hat Q \hat n_\theta \hat Q^{-1} = \hat n_\theta -1$, and therefore it commutes with the Hamiltonian $[H_{bos},\hat Q]=0$.

For a system of sufficient large length $L \gg 1$, the second term in Eq.~\eqref{zeromodeha} dominates and $e^{i \hat  \theta_0} = e^{2 i \pi \hat n_\theta/N}$ commutes with the Hamiltonian  $H_{bos}$ (it also commutes at finite length when $K \to \infty$ which is the case for the exact solution of Sec.~\ref{sec:exact}). The pair of operators 
$\hat \gamma_1$ and $\hat \gamma_2$ given by (also Eq.~\eqref{paraf} in the main text),
\begin{equation}\label{comm2}
  \hat \gamma_1 = e^{i \hat  \theta_0}, \qquad \qquad \hat \gamma_2 = \hat \gamma_1 e^{\frac{2 i \pi  (\hat N_e-1/2)}{N}}, \qquad \qquad \hat Q = \omega^{1/2}  \hat \gamma_1^\dagger \hat \gamma_2,
\end{equation}
commutes with the Hamiltonian, $[\hat \gamma_{1/2},H_{bos}]=0$, and satisfies commutation relations (see main text) characterizing them as parafermions.

The expressions for the parafermions can also be obtained from the analysis of the exact solution in Sec.~\ref{sec:paraf}. We bosonize the operator
\begin{equation}
  \tilde c_j  =  \frac{1}{\sqrt{2 \pi}} e^{i \hat \theta(j a_0) },
\end{equation}
which only involves the field $\hat \theta (x)$ due to the attached Jordan-Wigner string. We then make use of the expansion~\eqref{mode-exp2} in Eq.~\eqref{comm} and neglect the modes $\hat a_k$ by assuming proximity to the ground state. The result coincides with Eq.~\eqref{comm2} which strongly points towards to the fact that these parafermionic modes are non-local.

\subsection{$8 \pi$-periodic spectral flow dual to the fractional Josephson effect}

Electron-electron interactions in a quantum spin Hall edge state have been shown~\cite{zhang2014,orth2015} to mediate a fractional Josephson effect with $8 \pi$ periodicity protected by time-reversal symmetry and fermion parity. As we now discuss, our model of one-dimensional fermions with multiplet pairing is dual to model of Ref.~\onlinecite{zhang2014} and therefore predicts a similar ground state anomaly.

We focus on the case  $N=4$ for simplicity.
At low energy (short-distance variations $\sim 1/k_F$ are discarded), the model exhibits an emergent antiunitary symmetry $\Xi$
\begin{equation}
 i \to  -i \qquad \hat \psi_R (x) \to \hat \psi_L^\dagger (x) \qquad \hat \psi_L (x) \to  - \hat \psi_R^\dagger (x) \qquad \Xi^2= -1
\end{equation}
commuting with the Hamiltonian $[\Xi,H]=0$. In terms of bosonic fields, the symmetry acts as $\hat \theta \to  \hat \theta + \pi/2$, $\hat \phi \to  - \hat \phi - \pi/2$. Exchanging the transformations on $\hat \theta$ and $\hat \varphi$ precisely recovers time-reversal symmetry, namely $\psi_R \to \psi_L$, $\psi_L \to -\psi_R$.   Whereas a local Zeeman term $\sim \cos 2 \hat \phi$ breaks time-reversal symmetry in Ref.~\onlinecite{zhang2014}, here the symmetry $\Xi$ is broken by a pairing term $\hat c_j  \hat c_{j+1} \sim \psi_R (x) \psi_L (x) \sim \cos 2 \hat \theta$ distinguishing the states  $| \theta_m \rangle$.

To reveal the ground state anomaly dual to the $8 \pi$ Josephson effect, we sandwich the multiplet pairing model (Eq.~\eqref{hp} in the main text) between two band insulators with commensurate potentials $\cos (2 k_F x + \varphi_{1/2})$. The phase $\varphi_1$ ($\varphi_2$) parametrizes the positions of the potential minima in the insulator located on the left (right). We set $\varphi_2=0$ for simplicity. The corresponding bosonized form of this backscattering term is $\psi_R^\dagger \psi_L e^{i \varphi_{1/2}} + h.c. \sim \cos(2 \hat \phi - \varphi_{1/2})$. The commensurate potentials induce backscattering of single electrons transferred between the right-moving and left-moving channels with the phases $\varphi_{1/2}$.

The setup thus alternates two regions dominated by $\cos(2 \hat \phi)$ with a central one gapped by $\cos(4 \hat \theta)$ and is thus dual to the model of Refs.~\cite{zhang2014,orth2015}. It is known~\cite{qi2008} that a magnetic domain wall in a quantum spin Hall edge state binds half an electronic charge $e/2$ and that the same charge $e/2$ is also pumped when the magnetization is reversed. This is due to the pinning of the field $\hat \phi$ by the $\cos(2 \hat \phi)$ term which shifts its minima under these processes. In our setup, a tunneling event between the states $ \theta_m $ and  $\theta_{m+1}$ involves a change in the conjugated field $\Delta \theta =  \pi /2$ corresponding to the transfer of a fractional charge  $\Delta \theta/2 \pi = 1/4$ between the right-moving and left-moving channels. Such instanton tunneling is accompanied by the phase term $e^{i \varphi_{1}/4}$ and the Hamiltonian governing the low-energy subspace is given by ($m=4$ and $m=0$ are identified)~\cite{zhang2014}
\begin{equation}
  H = \Gamma \sum_{m=0}^3 e^{i \varphi_{1}/4} | \theta_m \rangle \langle \theta_{m+1} | + h.c.
\end{equation}
where the energy scale $\Gamma$ for instanton tunneling is exponentially small with the size $L$. The resulting spectral flow is $8 \pi$ periodic in $\varphi_{1}$. The energy crossings are protected by $\mathbb{Z}_{4}$ symmetry, {\it i.e.} each eigenstate has a well-defined number of fermion modulo $4$. Similarly to its dual model~\cite{zhang2014}, the crossings at $\varphi_1=0 \, [2 \pi]$ are also protected by the symmetry $\Xi$ while the crossings at $\varphi_1= \pi \, [2 \pi]$ are protected by standard fermion number parity. The former are splitted in the presence of a local pairing term $\sim \cos 2 \hat \theta$ breaking  $\Xi$ whereas the latter are topologically protected by the edge Majorana zero modes.

\subsection{Majorana edge modes}

In fractional quantum Hall states with proximity s-wave superconductivity, parafermionic operators have been explicitely constructed by Clarke, Alicea and Shtengel~\cite{clarke2013} and rigorously shown to be localized at edges. We adapt below their analysis to our model and show that edge Majorana fermions rather than edge parafermions  thus emerge. We first add two regions of small size $\ell \ll L$ where $\Delta =0$. They are localed at the left $- \ell <x<0$ and right  $L <x<L+\ell$ sides of the system. In the left region, the bosonic fields $\hat \varphi_{R/L} (x) = \hat \varphi (x) \pm \theta (x)$ are given by
\begin{equation}\label{mode-exp3}
  \hat \varphi_{R/L} (x)   =  \pm  \frac{2 \pi \hat n_\theta}{N}  \pm \sqrt{2} \sum_{q=0}^{\infty}  \frac{\left( e^{\pm i \lambda_q (x)} \hat b_q + h.c. \right)}{\sqrt{2 q + 1}} \qquad \qquad \begin{cases} \hat \varphi_{L} (- \ell) =  - \hat \varphi_{R} ( - \ell) \\[1mm] \hat \varphi_{L} (0) =  \hat \varphi_{R} (0) - 4 \pi \hat n_\theta/N
  \end{cases}
\end{equation}
with $\lambda_q (x) = \frac{(2 q+1) \pi (x + \ell)}{2 \ell}$ and the bosonic modes $\hat b_q$ satisfying $[\hat b_q,\hat b_{q'}^\dagger] = \delta_{q,q'}$. Using the Hamiltonian~\eqref{bos} (with $\Delta=0$), the following identity can be derived (we consider $K=1$ for simplicity)
\begin{equation}
  [ H_{bos}, e^{i  \hat \varphi_{R/L} (x) } ] = \pm i v_F \partial_x e^{i  \hat \varphi_{R/L} (x) },
\end{equation}
such that the operator
\begin{equation}
  \hat \beta_1 = e^{\frac{2 i \pi \hat n_\theta}{N}} \int_{- \ell}^0 d x \left[  e^{-\frac{2 i \pi \hat n_\theta}{N}} e^{i \hat \varphi_{R} (x)} +  e^{\frac{2 i \pi \hat n_\theta}{N}} e^{i \hat \varphi_{L} (x)} + h.c. \right]
\end{equation}
is a zero-mode commuting with the Hamiltonian $H_{bos}$. Noting that  $\hat \varphi_{L} (0) =  \hat \varphi_{R} (0) - 4 \pi \hat n_\theta/N$, we observe that the expression of  $\hat \beta_1$ involves only the local fermionic fields $e^{\pm i  \hat \varphi_{R/L} (x,0)}$ localized inside the small region left to the system. At low energy and for $\ell \to 0$, the modes $\hat b_q$ are essentially gapped~\cite{clarke2013} and the boundary mode simplifies as $\hat \beta_1 \sim e^{\frac{2 i \pi \hat n_\theta}{N}}$ as given by Eq.~\eqref{majoranas} in the main text. A similar construction gives the right boundary operator $\hat \beta_2$ also given Eq.~\eqref{majoranas}.

\section{Absence of edge parafermions in one dimensional fermionic models}

In this section we present details for the demonstration 
that parafermionic statistics cannot be obtained from spatially separated operators
in one-dimensional fermionic systems.

 Let us first introduce $2L$ Majorana fermions $\hat \alpha_j$ for the $L$ sites of model,
 satisfying the usual anticommutation relations:
 \begin{equation}
\{\hat \alpha_j,\hat \alpha_\ell\} = 2 \delta_{j,\ell},\qquad \hat \alpha_j^\dagger  = \hat \alpha_j,
\end{equation}
with $j=1,..,2L$. A complete basis for the full Hilbert space $\mathcal{H}$, of dimension $2^L$, can be constructed 
 in terms of products of such Majorana fermions, as follows \cite{Prosen2008,Jaffe2015},
 \begin{equation}
 \hat \Gamma_{\vec{j}} \equiv \hat \alpha_{j_1} \hat \alpha_{j_2} ... \hat \alpha_{j_m},
 \end{equation}
 where $\vec{j} = (j_1, j_2, ..., j_m)$ is an ordered vector of size $m$
 parametrizing a list of occupied modes, with $j_i = 1,...,L$ and $m=0,...,L$. 
 For the special case $m=0$ ones defines $\hat \Gamma_{\vec{j}} \equiv \mathbb{I}$.
The operators $\hat \Gamma_{\vec{j}}$ are linearly independent
and any operator $\hat O$ in the Hilbert space has the decomposition,
 \begin{equation} \label{eq.supp:operator.expansion}
\hat O = \sum_{ \{\vec{j} \mid j_i \in S(O) \} } f_{\vec{j}} \,\hat \Gamma_{\vec{j}}
 \end{equation}
 where $S(O) = (s_1,s_2,...,s_{L_O})$  denotes the support of the operator along the sites of model,
  with $s_i=1,..,L$ representing the sites the operator occupies, and $L_O = 1,...,L$ its effective length.
 
 Consider now two operators $O_A$ and $O_B$, with local supports $S(A)$ and $S(B)$ in exclusive regions
  of space: $S(A) \cap S(B) = \emptyset$. Using the decomposition of Eq.\eqref{eq.supp:operator.expansion}, 
  the commutation relation $\hat O_A \hat O_B + \varepsilon \hat O_B \hat O_A $, with $\varepsilon$ an \textit{a priori} arbitrary phase, 
  can be written as,
 \begin{eqnarray}
 \hat O_A \hat O_B + \varepsilon \hat O_B \hat O_A &=& \sum_{ \{\vec{j}_A \mid j_{A_i} \in S(A)\} \atop \{\vec{j}_B \mid j_{B_i} \in S(B)\}}
 f_{\vec{j}_A} f_{\vec{j}_B} \, (\hat \Gamma_{\vec{j}_A}\hat  \Gamma_{\vec{j}_B } + \varepsilon \hat  \Gamma_{\vec{j}_B} \hat \Gamma_{\vec{j}_A }) \\
 &=&  \sum_{ \{\vec{j}_A \mid j_{A_i} \in S(A)\} \atop \{\vec{j}_B \mid j_{B_i} \in S(B)\}}
 f_{\vec{j}_A}f_{\vec{j}_B} \, \hat \Gamma_{\vec{j}_A}\hat  \Gamma_{\vec{j}_B } (1 +\varepsilon  \phi_{\vec{j}_A,\vec{j}_B} )\\
 &=&  \sum_{ \{\vec{j}_A \mid j_{A_i} \in S(A)\} \atop \{\vec{j}_B \mid j_{B_i} \in S(B)\}}
 f_{\vec{j}_A}f_{\vec{j}_B} \, \hat \Gamma_{ \vec{j}_{AB} \equiv \vec{j}_A\cup \vec{j}_B }  (1 +\varepsilon \phi_{\vec{j}_A,\vec{j}_B} ) \label{supp.eq:comm.rel3}
 \end{eqnarray}
where in the second line we used the commutation relation $\hat \Gamma_{\vec{j}_B} \hat \Gamma_{\vec{j}_A } = 
\phi_{\vec{j}_A,\vec{j}_B} \hat \Gamma_{\vec{j}_A} \hat \Gamma_{\vec{j}_B }$, with $\phi_{\vec{j}_A,\vec{j}_B}$ a phase 
which can assume values $\pm 1$, depending only on the length of the vectors $\vec{j}_{A}$ and $\vec{j}_{B}$.
 In the third line we used the fact that 
 $\hat \Gamma_{\vec{j}_{AB} \equiv \vec{j}_A\cup \vec{j}_B} =  \hat \Gamma_{\vec{j}_A}\hat  \Gamma_{\vec{j}_B } $,
 since the support of the operators $ \hat \Gamma_{\vec{j}_A}$ and $\hat  \Gamma_{\vec{j}_B } $ are in exclusive regions 
 of space with their respective vectors $\vec{j}_{A}$ and $\vec{j}_{B}$ not overlapping between each other.
 
 Notice that the operators $\hat \Gamma_{\vec{j}_{AB} }$ in the sum of Eq.\eqref{supp.eq:comm.rel3} are orthogonal to each other.
 Thus, in order to $\hat O_A$ and $\hat O_B$ satisfy the commutation relation $\hat O_A \hat O_B + \varepsilon \hat O_B \hat O_A =0$ 
 all of the terms in the sum 
 must vanish,
 \begin{equation}
 (1 +\varepsilon \phi_{\vec{j}_A,\vec{j}_B} ) = 0,\qquad \forall \vec{j}_A,\vec{j}_B,
 \end{equation}
 a condition that can only be satisfied if $\varepsilon = \pm 1$, corresponding in this way 
 to fermionic or bosonic commutation relations.

\bibliography{pairing}

\begin{thebibliography}{88}%
\makeatletter
\providecommand \@ifxundefined [1]{%
 \@ifx{#1\undefined}
}%
\providecommand \@ifnum [1]{%
 \ifnum #1\expandafter \@firstoftwo
 \else \expandafter \@secondoftwo
 \fi
}%
\providecommand \@ifx [1]{%
 \ifx #1\expandafter \@firstoftwo
 \else \expandafter \@secondoftwo
 \fi
}%
\providecommand \natexlab [1]{#1}%
\providecommand \enquote  [1]{``#1''}%
\providecommand \bibnamefont  [1]{#1}%
\providecommand \bibfnamefont [1]{#1}%
\providecommand \citenamefont [1]{#1}%
\providecommand \href@noop [0]{\@secondoftwo}%
\providecommand \href [0]{\begingroup \@sanitize@url \@href}%
\providecommand \@href[1]{\@@startlink{#1}\@@href}%
\providecommand \@@href[1]{\endgroup#1\@@endlink}%
\providecommand \@sanitize@url [0]{\catcode `\\12\catcode `\$12\catcode
  `\&12\catcode `\#12\catcode `\^12\catcode `\_12\catcode `\%12\relax}%
\providecommand \@@startlink[1]{}%
\providecommand \@@endlink[0]{}%
\providecommand \url  [0]{\begingroup\@sanitize@url \@url }%
\providecommand \@url [1]{\endgroup\@href {#1}{\urlprefix }}%
\providecommand \urlprefix  [0]{URL }%
\providecommand \Eprint [0]{\href }%
\providecommand \doibase [0]{http://dx.doi.org/}%
\providecommand \selectlanguage [0]{\@gobble}%
\providecommand \bibinfo  [0]{\@secondoftwo}%
\providecommand \bibfield  [0]{\@secondoftwo}%
\providecommand \translation [1]{[#1]}%
\providecommand \BibitemOpen [0]{}%
\providecommand \bibitemStop [0]{}%
\providecommand \bibitemNoStop [0]{.\EOS\space}%
\providecommand \EOS [0]{\spacefactor3000\relax}%
\providecommand \BibitemShut  [1]{\csname bibitem#1\endcsname}%
\let\auto@bib@innerbib\@empty
\bibitem [{\citenamefont {Kitaev}(2001)}]{Kitaev2001}%
  \BibitemOpen
  \bibfield  {author} {\bibinfo {author} {\bibfnamefont {A.}~\bibnamefont
  {Kitaev}},\ }\href {\doibase 10.1070/1063-7869/44/10S/S29} {\bibfield
  {journal} {\bibinfo  {journal} {Physics Uspekhi}\ }\textbf {\bibinfo {volume}
  {44}},\ \bibinfo {pages} {131} (\bibinfo {year} {2001})}\BibitemShut
  {NoStop}%
\bibitem [{\citenamefont {Alicea}(2012)}]{alicea2012}%
  \BibitemOpen
  \bibfield  {author} {\bibinfo {author} {\bibfnamefont {J.}~\bibnamefont
  {Alicea}},\ }\href@noop {} {\bibfield  {journal} {\bibinfo  {journal} {Rep.
  Prog. Phys. 75, 076501}\ } (\bibinfo {year} {2012})}\BibitemShut {NoStop}%
\bibitem [{\citenamefont {Beenakker}(2013)}]{Beenakker2013}%
  \BibitemOpen
  \bibfield  {author} {\bibinfo {author} {\bibfnamefont {C.}~\bibnamefont
  {Beenakker}},\ }\href@noop {} {\bibfield  {journal} {\bibinfo  {journal}
  {Annual Review of Condensed Matter Physics}\ }\textbf {\bibinfo {volume}
  {4}},\ \bibinfo {pages} {113} (\bibinfo {year} {2013})}\BibitemShut {NoStop}%
\bibitem [{\citenamefont {Guo}(2016)}]{Guo2016}%
  \BibitemOpen
  \bibfield  {author} {\bibinfo {author} {\bibfnamefont {H.-M.}\ \bibnamefont
  {Guo}},\ }\href {\doibase 10.1007/s11433-015-5773-5} {\bibfield  {journal}
  {\bibinfo  {journal} {Science China Physics, Mechanics {\&} Astronomy}\
  }\textbf {\bibinfo {volume} {59}},\ \bibinfo {pages} {637401} (\bibinfo
  {year} {2016})}\BibitemShut {NoStop}%
\bibitem [{\citenamefont {Lutchyn}\ \emph {et~al.}(2018)\citenamefont
  {Lutchyn}, \citenamefont {Bakkers}, \citenamefont {Kouwenhoven},
  \citenamefont {Krogstrup}, \citenamefont {Marcus},\ and\ \citenamefont
  {Oreg}}]{lutchyn2017realizing}%
  \BibitemOpen
  \bibfield  {author} {\bibinfo {author} {\bibfnamefont {R.~M.}\ \bibnamefont
  {Lutchyn}}, \bibinfo {author} {\bibfnamefont {E.~P. A.~M.}\ \bibnamefont
  {Bakkers}}, \bibinfo {author} {\bibfnamefont {L.~P.}\ \bibnamefont
  {Kouwenhoven}}, \bibinfo {author} {\bibfnamefont {P.}~\bibnamefont
  {Krogstrup}}, \bibinfo {author} {\bibfnamefont {C.~M.}\ \bibnamefont
  {Marcus}}, \ and\ \bibinfo {author} {\bibfnamefont {Y.}~\bibnamefont
  {Oreg}},\ }\href {\doibase 10.1038/s41578-018-0003-1} {\bibfield  {journal}
  {\bibinfo  {journal} {Nature Reviews Materials}\ }\textbf {\bibinfo {volume}
  {3}},\ \bibinfo {pages} {52} (\bibinfo {year} {2018})}\BibitemShut {NoStop}%
\bibitem [{\citenamefont {Oreg}\ \emph {et~al.}(2010)\citenamefont {Oreg},
  \citenamefont {Refael},\ and\ \citenamefont {von Oppen}}]{wire1}%
  \BibitemOpen
  \bibfield  {author} {\bibinfo {author} {\bibfnamefont {Y.}~\bibnamefont
  {Oreg}}, \bibinfo {author} {\bibfnamefont {G.}~\bibnamefont {Refael}}, \ and\
  \bibinfo {author} {\bibfnamefont {F.}~\bibnamefont {von Oppen}},\ }\href
  {http://link.aps.org/doi/10.1103/PhysRevLett.105.177002} {\bibfield
  {journal} {\bibinfo  {journal} {Phys. Rev. Lett. 105, 177002}\ } (\bibinfo
  {year} {2010})}\BibitemShut {NoStop}%
\bibitem [{\citenamefont {Lutchyn}\ \emph {et~al.}(2010)\citenamefont
  {Lutchyn}, \citenamefont {Sau},\ and\ \citenamefont {Das~Sarma}}]{wire2}%
  \BibitemOpen
  \bibfield  {author} {\bibinfo {author} {\bibfnamefont {R.}~\bibnamefont
  {Lutchyn}}, \bibinfo {author} {\bibfnamefont {J.}~\bibnamefont {Sau}}, \ and\
  \bibinfo {author} {\bibfnamefont {S.}~\bibnamefont {Das~Sarma}},\ }\href
  {http://link.aps.org/doi/10.1103/PhysRevLett.105.077001} {\bibfield
  {journal} {\bibinfo  {journal} {Phys. Rev. Lett. 105, 077001}\ } (\bibinfo
  {year} {2010})}\BibitemShut {NoStop}%
\bibitem [{\citenamefont {Mourik}\ \emph {et~al.}(2012)\citenamefont {Mourik},
  \citenamefont {Zuo}, \citenamefont {Frolov}, \citenamefont {Plissard},
  \citenamefont {Bakkers},\ and\ \citenamefont {Kouwenhoven.}}]{mourik2012}%
  \BibitemOpen
  \bibfield  {author} {\bibinfo {author} {\bibfnamefont {V.}~\bibnamefont
  {Mourik}}, \bibinfo {author} {\bibfnamefont {K.}~\bibnamefont {Zuo}},
  \bibinfo {author} {\bibfnamefont {S.}~\bibnamefont {Frolov}}, \bibinfo
  {author} {\bibfnamefont {S.}~\bibnamefont {Plissard}}, \bibinfo {author}
  {\bibfnamefont {E.}~\bibnamefont {Bakkers}}, \ and\ \bibinfo {author}
  {\bibfnamefont {L.}~\bibnamefont {Kouwenhoven.}},\ }\href@noop {} {\bibfield
  {journal} {\bibinfo  {journal} {Science 336, 1003}\ } (\bibinfo {year}
  {2012})}\BibitemShut {NoStop}%
\bibitem [{\citenamefont {{Albrecht}}\ \emph {et~al.}(2016)\citenamefont
  {{Albrecht}}, \citenamefont {{Higginbotham}}, \citenamefont {{Madsen}},
  \citenamefont {{Kuemmeth}}, \citenamefont {{Jespersen}}, \citenamefont
  {{Nyg{\aa}rd}}, \citenamefont {{Krogstrup}},\ and\ \citenamefont
  {{Marcus}}}]{Albrecht2016}%
  \BibitemOpen
  \bibfield  {author} {\bibinfo {author} {\bibfnamefont {S.~M.}\ \bibnamefont
  {{Albrecht}}}, \bibinfo {author} {\bibfnamefont {A.~P.}\ \bibnamefont
  {{Higginbotham}}}, \bibinfo {author} {\bibfnamefont {M.}~\bibnamefont
  {{Madsen}}}, \bibinfo {author} {\bibfnamefont {F.}~\bibnamefont
  {{Kuemmeth}}}, \bibinfo {author} {\bibfnamefont {T.~S.}\ \bibnamefont
  {{Jespersen}}}, \bibinfo {author} {\bibfnamefont {J.}~\bibnamefont
  {{Nyg{\aa}rd}}}, \bibinfo {author} {\bibfnamefont {P.}~\bibnamefont
  {{Krogstrup}}}, \ and\ \bibinfo {author} {\bibfnamefont {C.~M.}\ \bibnamefont
  {{Marcus}}},\ }\href
  {http://www.nature.com/nature/journal/v531/n7593/full/nature17162.html}
  {\bibfield  {journal} {\bibinfo  {journal} {\nat}\ }\textbf {\bibinfo
  {volume} {531}},\ \bibinfo {pages} {206} (\bibinfo {year}
  {2016})}\BibitemShut {NoStop}%
\bibitem [{\citenamefont {Deng}\ \emph {et~al.}(2016)\citenamefont {Deng},
  \citenamefont {Vaitiekenas}, \citenamefont {Hansen}, \citenamefont {Danon},
  \citenamefont {Leijnse}, \citenamefont {Flensberg}, \citenamefont {Nyg{\r
  a}rd}, \citenamefont {Krogstrup},\ and\ \citenamefont {Marcus}}]{Deng1557}%
  \BibitemOpen
  \bibfield  {author} {\bibinfo {author} {\bibfnamefont {M.~T.}\ \bibnamefont
  {Deng}}, \bibinfo {author} {\bibfnamefont {S.}~\bibnamefont {Vaitiekenas}},
  \bibinfo {author} {\bibfnamefont {E.~B.}\ \bibnamefont {Hansen}}, \bibinfo
  {author} {\bibfnamefont {J.}~\bibnamefont {Danon}}, \bibinfo {author}
  {\bibfnamefont {M.}~\bibnamefont {Leijnse}}, \bibinfo {author} {\bibfnamefont
  {K.}~\bibnamefont {Flensberg}}, \bibinfo {author} {\bibfnamefont
  {J.}~\bibnamefont {Nyg{\r a}rd}}, \bibinfo {author} {\bibfnamefont
  {P.}~\bibnamefont {Krogstrup}}, \ and\ \bibinfo {author} {\bibfnamefont
  {C.~M.}\ \bibnamefont {Marcus}},\ }\href {\doibase 10.1126/science.aaf3961}
  {\bibfield  {journal} {\bibinfo  {journal} {Science}\ }\textbf {\bibinfo
  {volume} {354}},\ \bibinfo {pages} {1557} (\bibinfo {year}
  {2016})}\BibitemShut {NoStop}%
\bibitem [{\citenamefont {Chen}\ \emph {et~al.}(2017)\citenamefont {Chen},
  \citenamefont {Yu}, \citenamefont {Stenger}, \citenamefont {Hocevar},
  \citenamefont {Car}, \citenamefont {Plissard}, \citenamefont {Bakkers},
  \citenamefont {Stanescu},\ and\ \citenamefont {Frolov}}]{Chene1701476}%
  \BibitemOpen
  \bibfield  {author} {\bibinfo {author} {\bibfnamefont {J.}~\bibnamefont
  {Chen}}, \bibinfo {author} {\bibfnamefont {P.}~\bibnamefont {Yu}}, \bibinfo
  {author} {\bibfnamefont {J.}~\bibnamefont {Stenger}}, \bibinfo {author}
  {\bibfnamefont {M.}~\bibnamefont {Hocevar}}, \bibinfo {author} {\bibfnamefont
  {D.}~\bibnamefont {Car}}, \bibinfo {author} {\bibfnamefont {S.~R.}\
  \bibnamefont {Plissard}}, \bibinfo {author} {\bibfnamefont {E.~P. A.~M.}\
  \bibnamefont {Bakkers}}, \bibinfo {author} {\bibfnamefont {T.~D.}\
  \bibnamefont {Stanescu}}, \ and\ \bibinfo {author} {\bibfnamefont {S.~M.}\
  \bibnamefont {Frolov}},\ }\href {\doibase 10.1126/sciadv.1701476} {\bibfield
  {journal} {\bibinfo  {journal} {Science Advances}\ }\textbf {\bibinfo
  {volume} {3}} (\bibinfo {year} {2017}),\ 10.1126/sciadv.1701476}\BibitemShut
  {NoStop}%
\bibitem [{\citenamefont {Choy}\ \emph {et~al.}(2011)\citenamefont {Choy},
  \citenamefont {Edge}, \citenamefont {Akhmerov},\ and\ \citenamefont
  {Beenakker}}]{choy2011}%
  \BibitemOpen
  \bibfield  {author} {\bibinfo {author} {\bibfnamefont {T.-P.}\ \bibnamefont
  {Choy}}, \bibinfo {author} {\bibfnamefont {J.~M.}\ \bibnamefont {Edge}},
  \bibinfo {author} {\bibfnamefont {A.~R.}\ \bibnamefont {Akhmerov}}, \ and\
  \bibinfo {author} {\bibfnamefont {C.~W.~J.}\ \bibnamefont {Beenakker}},\
  }\href {\doibase 10.1103/PhysRevB.84.195442} {\bibfield  {journal} {\bibinfo
  {journal} {Phys. Rev. B}\ }\textbf {\bibinfo {volume} {84}},\ \bibinfo
  {pages} {195442} (\bibinfo {year} {2011})}\BibitemShut {NoStop}%
\bibitem [{\citenamefont {Nadj-Perge}\ \emph {et~al.}(2013)\citenamefont
  {Nadj-Perge}, \citenamefont {Drozdov}, \citenamefont {Bernevig},\ and\
  \citenamefont {Yazdani}}]{NadjPerge2013}%
  \BibitemOpen
  \bibfield  {author} {\bibinfo {author} {\bibfnamefont {S.}~\bibnamefont
  {Nadj-Perge}}, \bibinfo {author} {\bibfnamefont {I.~K.}\ \bibnamefont
  {Drozdov}}, \bibinfo {author} {\bibfnamefont {B.~A.}\ \bibnamefont
  {Bernevig}}, \ and\ \bibinfo {author} {\bibfnamefont {A.}~\bibnamefont
  {Yazdani}},\ }\href {\doibase 10.1103/PhysRevB.88.020407} {\bibfield
  {journal} {\bibinfo  {journal} {Phys. Rev. B}\ }\textbf {\bibinfo {volume}
  {88}},\ \bibinfo {pages} {020407} (\bibinfo {year} {2013})}\BibitemShut
  {NoStop}%
\bibitem [{\citenamefont {Braunecker}\ and\ \citenamefont
  {Simon}(2013)}]{Braunecker2013}%
  \BibitemOpen
  \bibfield  {author} {\bibinfo {author} {\bibfnamefont {B.}~\bibnamefont
  {Braunecker}}\ and\ \bibinfo {author} {\bibfnamefont {P.}~\bibnamefont
  {Simon}},\ }\href {\doibase 10.1103/PhysRevLett.111.147202} {\bibfield
  {journal} {\bibinfo  {journal} {Phys. Rev. Lett.}\ }\textbf {\bibinfo
  {volume} {111}},\ \bibinfo {pages} {147202} (\bibinfo {year}
  {2013})}\BibitemShut {NoStop}%
\bibitem [{\citenamefont {Vazifeh}\ and\ \citenamefont
  {Franz}(2013)}]{Vazifeh2013}%
  \BibitemOpen
  \bibfield  {author} {\bibinfo {author} {\bibfnamefont {M.~M.}\ \bibnamefont
  {Vazifeh}}\ and\ \bibinfo {author} {\bibfnamefont {M.}~\bibnamefont
  {Franz}},\ }\href {\doibase 10.1103/PhysRevLett.111.206802} {\bibfield
  {journal} {\bibinfo  {journal} {Phys. Rev. Lett.}\ }\textbf {\bibinfo
  {volume} {111}},\ \bibinfo {pages} {206802} (\bibinfo {year}
  {2013})}\BibitemShut {NoStop}%
\bibitem [{\citenamefont {Klinovaja}\ \emph {et~al.}(2013)\citenamefont
  {Klinovaja}, \citenamefont {Stano}, \citenamefont {Yazdani},\ and\
  \citenamefont {Loss}}]{klinovaja2013}%
  \BibitemOpen
  \bibfield  {author} {\bibinfo {author} {\bibfnamefont {J.}~\bibnamefont
  {Klinovaja}}, \bibinfo {author} {\bibfnamefont {P.}~\bibnamefont {Stano}},
  \bibinfo {author} {\bibfnamefont {A.}~\bibnamefont {Yazdani}}, \ and\
  \bibinfo {author} {\bibfnamefont {D.}~\bibnamefont {Loss}},\ }\href {\doibase
  10.1103/PhysRevLett.111.186805} {\bibfield  {journal} {\bibinfo  {journal}
  {Phys. Rev. Lett.}\ }\textbf {\bibinfo {volume} {111}},\ \bibinfo {pages}
  {186805} (\bibinfo {year} {2013})}\BibitemShut {NoStop}%
\bibitem [{\citenamefont {Nadj-Perge}\ \emph {et~al.}(2014)\citenamefont
  {Nadj-Perge}, \citenamefont {Drozdov}, \citenamefont {Li}, \citenamefont
  {Chen}, \citenamefont {Jeon}, \citenamefont {Seo}, \citenamefont {MacDonald},
  \citenamefont {Bernevig},\ and\ \citenamefont {Yazdani}}]{yazdani2014}%
  \BibitemOpen
  \bibfield  {author} {\bibinfo {author} {\bibfnamefont {S.}~\bibnamefont
  {Nadj-Perge}}, \bibinfo {author} {\bibfnamefont {I.~K.}\ \bibnamefont
  {Drozdov}}, \bibinfo {author} {\bibfnamefont {J.}~\bibnamefont {Li}},
  \bibinfo {author} {\bibfnamefont {H.}~\bibnamefont {Chen}}, \bibinfo {author}
  {\bibfnamefont {S.}~\bibnamefont {Jeon}}, \bibinfo {author} {\bibfnamefont
  {J.}~\bibnamefont {Seo}}, \bibinfo {author} {\bibfnamefont {A.~H.}\
  \bibnamefont {MacDonald}}, \bibinfo {author} {\bibfnamefont {B.~A.}\
  \bibnamefont {Bernevig}}, \ and\ \bibinfo {author} {\bibfnamefont
  {A.}~\bibnamefont {Yazdani}},\ }\href@noop {} {\bibfield  {journal} {\bibinfo
   {journal} {Science}\ }\textbf {\bibinfo {volume} {346}},\ \bibinfo {pages}
  {602} (\bibinfo {year} {2014})}\BibitemShut {NoStop}%
\bibitem [{\citenamefont {Ruby}\ \emph {et~al.}(2015)\citenamefont {Ruby},
  \citenamefont {Pientka}, \citenamefont {Peng}, \citenamefont {von Oppen},
  \citenamefont {Heinrich},\ and\ \citenamefont {Franke}}]{ruby2015}%
  \BibitemOpen
  \bibfield  {author} {\bibinfo {author} {\bibfnamefont {M.}~\bibnamefont
  {Ruby}}, \bibinfo {author} {\bibfnamefont {F.}~\bibnamefont {Pientka}},
  \bibinfo {author} {\bibfnamefont {Y.}~\bibnamefont {Peng}}, \bibinfo {author}
  {\bibfnamefont {F.}~\bibnamefont {von Oppen}}, \bibinfo {author}
  {\bibfnamefont {B.~W.}\ \bibnamefont {Heinrich}}, \ and\ \bibinfo {author}
  {\bibfnamefont {K.~J.}\ \bibnamefont {Franke}},\ }\href {\doibase
  10.1103/PhysRevLett.115.197204} {\bibfield  {journal} {\bibinfo  {journal}
  {Phys. Rev. Lett.}\ }\textbf {\bibinfo {volume} {115}},\ \bibinfo {pages}
  {197204} (\bibinfo {year} {2015})}\BibitemShut {NoStop}%
\bibitem [{\citenamefont {Pawlak}\ \emph {et~al.}(2016)\citenamefont {Pawlak},
  \citenamefont {Kisiel}, \citenamefont {Klinovaja}, \citenamefont {Meier},
  \citenamefont {Kawai}, \citenamefont {Glatzel}, \citenamefont {Loss},\ and\
  \citenamefont {Meyer}}]{pawlak2016}%
  \BibitemOpen
  \bibfield  {author} {\bibinfo {author} {\bibfnamefont {R.}~\bibnamefont
  {Pawlak}}, \bibinfo {author} {\bibfnamefont {M.}~\bibnamefont {Kisiel}},
  \bibinfo {author} {\bibfnamefont {J.}~\bibnamefont {Klinovaja}}, \bibinfo
  {author} {\bibfnamefont {T.}~\bibnamefont {Meier}}, \bibinfo {author}
  {\bibfnamefont {S.}~\bibnamefont {Kawai}}, \bibinfo {author} {\bibfnamefont
  {T.}~\bibnamefont {Glatzel}}, \bibinfo {author} {\bibfnamefont
  {D.}~\bibnamefont {Loss}}, \ and\ \bibinfo {author} {\bibfnamefont
  {E.}~\bibnamefont {Meyer}},\ }\href@noop {} {\bibfield  {journal} {\bibinfo
  {journal} {NPJ Quantum Information}\ }\textbf {\bibinfo {volume} {2}},\
  \bibinfo {pages} {16035} (\bibinfo {year} {2016})}\BibitemShut {NoStop}%
\bibitem [{\citenamefont {Feldman}\ \emph {et~al.}(2017)\citenamefont
  {Feldman}, \citenamefont {Randeria}, \citenamefont {Li}, \citenamefont
  {Jeon}, \citenamefont {Xie}, \citenamefont {Wang}, \citenamefont {Drozdov},
  \citenamefont {Bernevig},\ and\ \citenamefont {Yazdani}}]{feldman2017}%
  \BibitemOpen
  \bibfield  {author} {\bibinfo {author} {\bibfnamefont {B.~E.}\ \bibnamefont
  {Feldman}}, \bibinfo {author} {\bibfnamefont {M.~T.}\ \bibnamefont
  {Randeria}}, \bibinfo {author} {\bibfnamefont {J.}~\bibnamefont {Li}},
  \bibinfo {author} {\bibfnamefont {S.}~\bibnamefont {Jeon}}, \bibinfo {author}
  {\bibfnamefont {Y.}~\bibnamefont {Xie}}, \bibinfo {author} {\bibfnamefont
  {Z.}~\bibnamefont {Wang}}, \bibinfo {author} {\bibfnamefont {I.~K.}\
  \bibnamefont {Drozdov}}, \bibinfo {author} {\bibfnamefont {B.~A.}\
  \bibnamefont {Bernevig}}, \ and\ \bibinfo {author} {\bibfnamefont
  {A.}~\bibnamefont {Yazdani}},\ }\href
  {http://www.nature.com/nphys/journal/v13/n3/full/nphys3947.html} {\bibfield
  {journal} {\bibinfo  {journal} {Nature Physics}\ }\textbf {\bibinfo {volume}
  {13}},\ \bibinfo {pages} {286} (\bibinfo {year} {2017})}\BibitemShut
  {NoStop}%
\bibitem [{\citenamefont {Nayak}\ \emph {et~al.}(2008)\citenamefont {Nayak},
  \citenamefont {Simon}, \citenamefont {Stern}, \citenamefont {Freedman},\ and\
  \citenamefont {Das~Sarma}}]{nayak2008}%
  \BibitemOpen
  \bibfield  {author} {\bibinfo {author} {\bibfnamefont {C.}~\bibnamefont
  {Nayak}}, \bibinfo {author} {\bibfnamefont {S.~H.}\ \bibnamefont {Simon}},
  \bibinfo {author} {\bibfnamefont {A.}~\bibnamefont {Stern}}, \bibinfo
  {author} {\bibfnamefont {M.}~\bibnamefont {Freedman}}, \ and\ \bibinfo
  {author} {\bibfnamefont {S.}~\bibnamefont {Das~Sarma}},\ }\href {\doibase
  10.1103/RevModPhys.80.1083} {\bibfield  {journal} {\bibinfo  {journal} {Rev.
  Mod. Phys.}\ }\textbf {\bibinfo {volume} {80}},\ \bibinfo {pages} {1083}
  (\bibinfo {year} {2008})}\BibitemShut {NoStop}%
\bibitem [{\citenamefont {Kitaev}(2003)}]{kitaev2003}%
  \BibitemOpen
  \bibfield  {author} {\bibinfo {author} {\bibfnamefont {A.}~\bibnamefont
  {Kitaev}},\ }\href {\doibase https://doi.org/10.1016/S0003-4916(02)00018-0}
  {\bibfield  {journal} {\bibinfo  {journal} {Annals of Physics}\ }\textbf
  {\bibinfo {volume} {303}},\ \bibinfo {pages} {2 } (\bibinfo {year}
  {2003})}\BibitemShut {NoStop}%
\bibitem [{\citenamefont {Aasen}\ \emph {et~al.}(2016)\citenamefont {Aasen},
  \citenamefont {Hell}, \citenamefont {Mishmash}, \citenamefont {Higginbotham},
  \citenamefont {Danon}, \citenamefont {Leijnse}, \citenamefont {Jespersen},
  \citenamefont {Folk}, \citenamefont {Marcus}, \citenamefont {Flensberg},\
  and\ \citenamefont {Alicea}}]{aasen2016}%
  \BibitemOpen
  \bibfield  {author} {\bibinfo {author} {\bibfnamefont {D.}~\bibnamefont
  {Aasen}}, \bibinfo {author} {\bibfnamefont {M.}~\bibnamefont {Hell}},
  \bibinfo {author} {\bibfnamefont {R.~V.}\ \bibnamefont {Mishmash}}, \bibinfo
  {author} {\bibfnamefont {A.}~\bibnamefont {Higginbotham}}, \bibinfo {author}
  {\bibfnamefont {J.}~\bibnamefont {Danon}}, \bibinfo {author} {\bibfnamefont
  {M.}~\bibnamefont {Leijnse}}, \bibinfo {author} {\bibfnamefont {T.~S.}\
  \bibnamefont {Jespersen}}, \bibinfo {author} {\bibfnamefont {J.~A.}\
  \bibnamefont {Folk}}, \bibinfo {author} {\bibfnamefont {C.~M.}\ \bibnamefont
  {Marcus}}, \bibinfo {author} {\bibfnamefont {K.}~\bibnamefont {Flensberg}}, \
  and\ \bibinfo {author} {\bibfnamefont {J.}~\bibnamefont {Alicea}},\ }\href
  {\doibase 10.1103/PhysRevX.6.031016} {\bibfield  {journal} {\bibinfo
  {journal} {Phys. Rev. X}\ }\textbf {\bibinfo {volume} {6}},\ \bibinfo {pages}
  {031016} (\bibinfo {year} {2016})}\BibitemShut {NoStop}%
\bibitem [{\citenamefont {Fendley}(2012)}]{fendley2012}%
  \BibitemOpen
  \bibfield  {author} {\bibinfo {author} {\bibfnamefont {P.}~\bibnamefont
  {Fendley}},\ }\href {http://stacks.iop.org/1742-5468/2012/i=11/a=P11020}
  {\bibfield  {journal} {\bibinfo  {journal} {Journal of Statistical Mechanics:
  Theory and Experiment}\ }\textbf {\bibinfo {volume} {2012}},\ \bibinfo
  {pages} {P11020} (\bibinfo {year} {2012})}\BibitemShut {NoStop}%
\bibitem [{\citenamefont {Burrello}\ \emph {et~al.}(2013)\citenamefont
  {Burrello}, \citenamefont {van Heck},\ and\ \citenamefont
  {Cobanera}}]{PhysRevB.87.195422}%
  \BibitemOpen
  \bibfield  {author} {\bibinfo {author} {\bibfnamefont {M.}~\bibnamefont
  {Burrello}}, \bibinfo {author} {\bibfnamefont {B.}~\bibnamefont {van Heck}},
  \ and\ \bibinfo {author} {\bibfnamefont {E.}~\bibnamefont {Cobanera}},\
  }\href {\doibase 10.1103/PhysRevB.87.195422} {\bibfield  {journal} {\bibinfo
  {journal} {Phys. Rev. B}\ }\textbf {\bibinfo {volume} {87}},\ \bibinfo
  {pages} {195422} (\bibinfo {year} {2013})}\BibitemShut {NoStop}%
\bibitem [{\citenamefont {Alicea}\ and\ \citenamefont
  {Fendley}(2016)}]{alicea2016}%
  \BibitemOpen
  \bibfield  {author} {\bibinfo {author} {\bibfnamefont {J.}~\bibnamefont
  {Alicea}}\ and\ \bibinfo {author} {\bibfnamefont {P.}~\bibnamefont
  {Fendley}},\ }\href@noop {} {\bibfield  {journal} {\bibinfo  {journal}
  {Annual Review of Condensed Matter Physics}\ }\textbf {\bibinfo {volume}
  {7}},\ \bibinfo {pages} {119} (\bibinfo {year} {2016})}\BibitemShut {NoStop}%
\bibitem [{\citenamefont {Fendley}(2014)}]{fendley2014}%
  \BibitemOpen
  \bibfield  {author} {\bibinfo {author} {\bibfnamefont {P.}~\bibnamefont
  {Fendley}},\ }\href {http://stacks.iop.org/1751-8121/47/i=7/a=075001}
  {\bibfield  {journal} {\bibinfo  {journal} {Journal of Physics A:
  Mathematical and Theoretical}\ }\textbf {\bibinfo {volume} {47}},\ \bibinfo
  {pages} {075001} (\bibinfo {year} {2014})}\BibitemShut {NoStop}%
\bibitem [{\citenamefont {Cobanera}\ and\ \citenamefont
  {Ortiz}(2014)}]{cobanera2014}%
  \BibitemOpen
  \bibfield  {author} {\bibinfo {author} {\bibfnamefont {E.}~\bibnamefont
  {Cobanera}}\ and\ \bibinfo {author} {\bibfnamefont {G.}~\bibnamefont
  {Ortiz}},\ }\href {\doibase 10.1103/PhysRevA.89.012328} {\bibfield  {journal}
  {\bibinfo  {journal} {Phys. Rev. A}\ }\textbf {\bibinfo {volume} {89}},\
  \bibinfo {pages} {012328} (\bibinfo {year} {2014})}\BibitemShut {NoStop}%
\bibitem [{\citenamefont {Milsted}\ \emph {et~al.}(2014)\citenamefont
  {Milsted}, \citenamefont {Cobanera}, \citenamefont {Burrello},\ and\
  \citenamefont {Ortiz}}]{PhysRevB.90.195101}%
  \BibitemOpen
  \bibfield  {author} {\bibinfo {author} {\bibfnamefont {A.}~\bibnamefont
  {Milsted}}, \bibinfo {author} {\bibfnamefont {E.}~\bibnamefont {Cobanera}},
  \bibinfo {author} {\bibfnamefont {M.}~\bibnamefont {Burrello}}, \ and\
  \bibinfo {author} {\bibfnamefont {G.}~\bibnamefont {Ortiz}},\ }\href
  {\doibase 10.1103/PhysRevB.90.195101} {\bibfield  {journal} {\bibinfo
  {journal} {Phys. Rev. B}\ }\textbf {\bibinfo {volume} {90}},\ \bibinfo
  {pages} {195101} (\bibinfo {year} {2014})}\BibitemShut {NoStop}%
\bibitem [{\citenamefont {Mong}\ \emph {et~al.}(2014)\citenamefont {Mong},
  \citenamefont {Clarke}, \citenamefont {Alicea}, \citenamefont {Lindner},\
  and\ \citenamefont {Fendley}}]{mong2014}%
  \BibitemOpen
  \bibfield  {author} {\bibinfo {author} {\bibfnamefont {R.~S.~K.}\
  \bibnamefont {Mong}}, \bibinfo {author} {\bibfnamefont {D.~J.}\ \bibnamefont
  {Clarke}}, \bibinfo {author} {\bibfnamefont {J.}~\bibnamefont {Alicea}},
  \bibinfo {author} {\bibfnamefont {N.~H.}\ \bibnamefont {Lindner}}, \ and\
  \bibinfo {author} {\bibfnamefont {P.}~\bibnamefont {Fendley}},\ }\href
  {http://stacks.iop.org/1751-8121/47/i=45/a=452001} {\bibfield  {journal}
  {\bibinfo  {journal} {Journal of Physics A: Mathematical and Theoretical}\
  }\textbf {\bibinfo {volume} {47}},\ \bibinfo {pages} {452001} (\bibinfo
  {year} {2014})}\BibitemShut {NoStop}%
\bibitem [{\citenamefont {Jermyn}\ \emph {et~al.}(2014)\citenamefont {Jermyn},
  \citenamefont {Mong}, \citenamefont {Alicea},\ and\ \citenamefont
  {Fendley}}]{jermyn2014}%
  \BibitemOpen
  \bibfield  {author} {\bibinfo {author} {\bibfnamefont {A.~S.}\ \bibnamefont
  {Jermyn}}, \bibinfo {author} {\bibfnamefont {R.~S.~K.}\ \bibnamefont {Mong}},
  \bibinfo {author} {\bibfnamefont {J.}~\bibnamefont {Alicea}}, \ and\ \bibinfo
  {author} {\bibfnamefont {P.}~\bibnamefont {Fendley}},\ }\href {\doibase
  10.1103/PhysRevB.90.165106} {\bibfield  {journal} {\bibinfo  {journal} {Phys.
  Rev. B}\ }\textbf {\bibinfo {volume} {90}},\ \bibinfo {pages} {165106}
  (\bibinfo {year} {2014})}\BibitemShut {NoStop}%
\bibitem [{\citenamefont {Zhuang}\ \emph {et~al.}(2015)\citenamefont {Zhuang},
  \citenamefont {Changlani}, \citenamefont {Tubman},\ and\ \citenamefont
  {Hughes}}]{zhuang2015}%
  \BibitemOpen
  \bibfield  {author} {\bibinfo {author} {\bibfnamefont {Y.}~\bibnamefont
  {Zhuang}}, \bibinfo {author} {\bibfnamefont {H.~J.}\ \bibnamefont
  {Changlani}}, \bibinfo {author} {\bibfnamefont {N.~M.}\ \bibnamefont
  {Tubman}}, \ and\ \bibinfo {author} {\bibfnamefont {T.~L.}\ \bibnamefont
  {Hughes}},\ }\href {\doibase 10.1103/PhysRevB.92.035154} {\bibfield
  {journal} {\bibinfo  {journal} {Phys. Rev. B}\ }\textbf {\bibinfo {volume}
  {92}},\ \bibinfo {pages} {035154} (\bibinfo {year} {2015})}\BibitemShut
  {NoStop}%
\bibitem [{\citenamefont {Stoudenmire}\ \emph {et~al.}(2015)\citenamefont
  {Stoudenmire}, \citenamefont {Clarke}, \citenamefont {Mong},\ and\
  \citenamefont {Alicea}}]{stoudenmire2015}%
  \BibitemOpen
  \bibfield  {author} {\bibinfo {author} {\bibfnamefont {E.~M.}\ \bibnamefont
  {Stoudenmire}}, \bibinfo {author} {\bibfnamefont {D.~J.}\ \bibnamefont
  {Clarke}}, \bibinfo {author} {\bibfnamefont {R.~S.~K.}\ \bibnamefont {Mong}},
  \ and\ \bibinfo {author} {\bibfnamefont {J.}~\bibnamefont {Alicea}},\ }\href
  {\doibase 10.1103/PhysRevB.91.235112} {\bibfield  {journal} {\bibinfo
  {journal} {Phys. Rev. B}\ }\textbf {\bibinfo {volume} {91}},\ \bibinfo
  {pages} {235112} (\bibinfo {year} {2015})}\BibitemShut {NoStop}%
\bibitem [{\citenamefont {Sreejith}\ \emph {et~al.}(2016)\citenamefont
  {Sreejith}, \citenamefont {Lazarides},\ and\ \citenamefont
  {Moessner}}]{Sreejith2016}%
  \BibitemOpen
  \bibfield  {author} {\bibinfo {author} {\bibfnamefont {G.~J.}\ \bibnamefont
  {Sreejith}}, \bibinfo {author} {\bibfnamefont {A.}~\bibnamefont {Lazarides}},
  \ and\ \bibinfo {author} {\bibfnamefont {R.}~\bibnamefont {Moessner}},\
  }\href {\doibase 10.1103/PhysRevB.94.045127} {\bibfield  {journal} {\bibinfo
  {journal} {Phys. Rev. B}\ }\textbf {\bibinfo {volume} {94}},\ \bibinfo
  {pages} {045127} (\bibinfo {year} {2016})}\BibitemShut {NoStop}%
\bibitem [{\citenamefont {Alexandradinata}\ \emph {et~al.}(2016)\citenamefont
  {Alexandradinata}, \citenamefont {Regnault}, \citenamefont {Fang},
  \citenamefont {Gilbert},\ and\ \citenamefont
  {Bernevig}}]{Alexandradinata2016}%
  \BibitemOpen
  \bibfield  {author} {\bibinfo {author} {\bibfnamefont {A.}~\bibnamefont
  {Alexandradinata}}, \bibinfo {author} {\bibfnamefont {N.}~\bibnamefont
  {Regnault}}, \bibinfo {author} {\bibfnamefont {C.}~\bibnamefont {Fang}},
  \bibinfo {author} {\bibfnamefont {M.~J.}\ \bibnamefont {Gilbert}}, \ and\
  \bibinfo {author} {\bibfnamefont {B.~A.}\ \bibnamefont {Bernevig}},\ }\href
  {\doibase 10.1103/PhysRevB.94.125103} {\bibfield  {journal} {\bibinfo
  {journal} {Phys. Rev. B}\ }\textbf {\bibinfo {volume} {94}},\ \bibinfo
  {pages} {125103} (\bibinfo {year} {2016})}\BibitemShut {NoStop}%
\bibitem [{\citenamefont {Chen}\ and\ \citenamefont
  {Burnell}(2016)}]{chen2016}%
  \BibitemOpen
  \bibfield  {author} {\bibinfo {author} {\bibfnamefont {C.}~\bibnamefont
  {Chen}}\ and\ \bibinfo {author} {\bibfnamefont {F.~J.}\ \bibnamefont
  {Burnell}},\ }\href {\doibase 10.1103/PhysRevLett.116.106405} {\bibfield
  {journal} {\bibinfo  {journal} {Phys. Rev. Lett.}\ }\textbf {\bibinfo
  {volume} {116}},\ \bibinfo {pages} {106405} (\bibinfo {year}
  {2016})}\BibitemShut {NoStop}%
\bibitem [{\citenamefont {Ebisu}\ \emph {et~al.}(2017)\citenamefont {Ebisu},
  \citenamefont {Sagi}, \citenamefont {Tanaka},\ and\ \citenamefont
  {Oreg}}]{hiromi2017}%
  \BibitemOpen
  \bibfield  {author} {\bibinfo {author} {\bibfnamefont {H.}~\bibnamefont
  {Ebisu}}, \bibinfo {author} {\bibfnamefont {E.}~\bibnamefont {Sagi}},
  \bibinfo {author} {\bibfnamefont {Y.}~\bibnamefont {Tanaka}}, \ and\ \bibinfo
  {author} {\bibfnamefont {Y.}~\bibnamefont {Oreg}},\ }\href {\doibase
  10.1103/PhysRevB.95.075111} {\bibfield  {journal} {\bibinfo  {journal} {Phys.
  Rev. B}\ }\textbf {\bibinfo {volume} {95}},\ \bibinfo {pages} {075111}
  (\bibinfo {year} {2017})}\BibitemShut {NoStop}%
\bibitem [{\citenamefont {Meidan}\ \emph {et~al.}(2017)\citenamefont {Meidan},
  \citenamefont {Berg},\ and\ \citenamefont {Stern}}]{meidan2017}%
  \BibitemOpen
  \bibfield  {author} {\bibinfo {author} {\bibfnamefont {D.}~\bibnamefont
  {Meidan}}, \bibinfo {author} {\bibfnamefont {E.}~\bibnamefont {Berg}}, \ and\
  \bibinfo {author} {\bibfnamefont {A.}~\bibnamefont {Stern}},\ }\href
  {\doibase 10.1103/PhysRevB.95.205104} {\bibfield  {journal} {\bibinfo
  {journal} {Phys. Rev. B}\ }\textbf {\bibinfo {volume} {95}},\ \bibinfo
  {pages} {205104} (\bibinfo {year} {2017})}\BibitemShut {NoStop}%
\bibitem [{\citenamefont {Xu}\ and\ \citenamefont {Zhang}(2017)}]{xu2017}%
  \BibitemOpen
  \bibfield  {author} {\bibinfo {author} {\bibfnamefont {W.-T.}\ \bibnamefont
  {Xu}}\ and\ \bibinfo {author} {\bibfnamefont {G.-M.}\ \bibnamefont {Zhang}},\
  }\href {\doibase 10.1103/PhysRevB.95.195122} {\bibfield  {journal} {\bibinfo
  {journal} {Phys. Rev. B}\ }\textbf {\bibinfo {volume} {95}},\ \bibinfo
  {pages} {195122} (\bibinfo {year} {2017})}\BibitemShut {NoStop}%
\bibitem [{\citenamefont {Moran}\ \emph {et~al.}(2017)\citenamefont {Moran},
  \citenamefont {Pellegrino}, \citenamefont {Slingerland},\ and\ \citenamefont
  {Kells}}]{moran2017}%
  \BibitemOpen
  \bibfield  {author} {\bibinfo {author} {\bibfnamefont {N.}~\bibnamefont
  {Moran}}, \bibinfo {author} {\bibfnamefont {D.}~\bibnamefont {Pellegrino}},
  \bibinfo {author} {\bibfnamefont {J.~K.}\ \bibnamefont {Slingerland}}, \ and\
  \bibinfo {author} {\bibfnamefont {G.}~\bibnamefont {Kells}},\ }\href
  {\doibase 10.1103/PhysRevB.95.235127} {\bibfield  {journal} {\bibinfo
  {journal} {Phys. Rev. B}\ }\textbf {\bibinfo {volume} {95}},\ \bibinfo
  {pages} {235127} (\bibinfo {year} {2017})}\BibitemShut {NoStop}%
\bibitem [{\citenamefont {Meichanetzidis}\ \emph {et~al.}(2018)\citenamefont
  {Meichanetzidis}, \citenamefont {Turner}, \citenamefont {Farjami},
  \citenamefont {Papi\ifmmode~\acute{c}\else \'{c}\fi{}},\ and\ \citenamefont
  {Pachos}}]{meichanetzidisfree}%
  \BibitemOpen
  \bibfield  {author} {\bibinfo {author} {\bibfnamefont {K.}~\bibnamefont
  {Meichanetzidis}}, \bibinfo {author} {\bibfnamefont {C.~J.}\ \bibnamefont
  {Turner}}, \bibinfo {author} {\bibfnamefont {A.}~\bibnamefont {Farjami}},
  \bibinfo {author} {\bibfnamefont {Z.}~\bibnamefont
  {Papi\ifmmode~\acute{c}\else \'{c}\fi{}}}, \ and\ \bibinfo {author}
  {\bibfnamefont {J.~K.}\ \bibnamefont {Pachos}},\ }\href {\doibase
  10.1103/PhysRevB.97.125104} {\bibfield  {journal} {\bibinfo  {journal} {Phys.
  Rev. B}\ }\textbf {\bibinfo {volume} {97}},\ \bibinfo {pages} {125104}
  (\bibinfo {year} {2018})}\BibitemShut {NoStop}%
\bibitem [{\citenamefont {Lindner}\ \emph {et~al.}(2012)\citenamefont
  {Lindner}, \citenamefont {Berg}, \citenamefont {Refael},\ and\ \citenamefont
  {Stern}}]{lindner2012}%
  \BibitemOpen
  \bibfield  {author} {\bibinfo {author} {\bibfnamefont {N.~H.}\ \bibnamefont
  {Lindner}}, \bibinfo {author} {\bibfnamefont {E.}~\bibnamefont {Berg}},
  \bibinfo {author} {\bibfnamefont {G.}~\bibnamefont {Refael}}, \ and\ \bibinfo
  {author} {\bibfnamefont {A.}~\bibnamefont {Stern}},\ }\href {\doibase
  10.1103/PhysRevX.2.041002} {\bibfield  {journal} {\bibinfo  {journal} {Phys.
  Rev. X}\ }\textbf {\bibinfo {volume} {2}},\ \bibinfo {pages} {041002}
  (\bibinfo {year} {2012})}\BibitemShut {NoStop}%
\bibitem [{\citenamefont {Cheng}(2012)}]{cheng2012}%
  \BibitemOpen
  \bibfield  {author} {\bibinfo {author} {\bibfnamefont {M.}~\bibnamefont
  {Cheng}},\ }\href {\doibase 10.1103/PhysRevB.86.195126} {\bibfield  {journal}
  {\bibinfo  {journal} {Phys. Rev. B}\ }\textbf {\bibinfo {volume} {86}},\
  \bibinfo {pages} {195126} (\bibinfo {year} {2012})}\BibitemShut {NoStop}%
\bibitem [{\citenamefont {Clarke}\ \emph {et~al.}(2013)\citenamefont {Clarke},
  \citenamefont {Alicea},\ and\ \citenamefont {Shtengel}}]{clarke2013}%
  \BibitemOpen
  \bibfield  {author} {\bibinfo {author} {\bibfnamefont {D.~J.}\ \bibnamefont
  {Clarke}}, \bibinfo {author} {\bibfnamefont {J.}~\bibnamefont {Alicea}}, \
  and\ \bibinfo {author} {\bibfnamefont {K.}~\bibnamefont {Shtengel}},\
  }\href@noop {} {\bibfield  {journal} {\bibinfo  {journal} {Nature
  communications}\ }\textbf {\bibinfo {volume} {4}},\ \bibinfo {pages} {1348}
  (\bibinfo {year} {2013})}\BibitemShut {NoStop}%
\bibitem [{\citenamefont {Vaezi}(2013)}]{vaezi2013}%
  \BibitemOpen
  \bibfield  {author} {\bibinfo {author} {\bibfnamefont {A.}~\bibnamefont
  {Vaezi}},\ }\href {\doibase 10.1103/PhysRevB.87.035132} {\bibfield  {journal}
  {\bibinfo  {journal} {Phys. Rev. B}\ }\textbf {\bibinfo {volume} {87}},\
  \bibinfo {pages} {035132} (\bibinfo {year} {2013})}\BibitemShut {NoStop}%
\bibitem [{\citenamefont {Motruk}\ \emph {et~al.}(2013)\citenamefont {Motruk},
  \citenamefont {Berg}, \citenamefont {Turner},\ and\ \citenamefont
  {Pollmann}}]{motruck2013}%
  \BibitemOpen
  \bibfield  {author} {\bibinfo {author} {\bibfnamefont {J.}~\bibnamefont
  {Motruk}}, \bibinfo {author} {\bibfnamefont {E.}~\bibnamefont {Berg}},
  \bibinfo {author} {\bibfnamefont {A.~M.}\ \bibnamefont {Turner}}, \ and\
  \bibinfo {author} {\bibfnamefont {F.}~\bibnamefont {Pollmann}},\ }\href
  {\doibase 10.1103/PhysRevB.88.085115} {\bibfield  {journal} {\bibinfo
  {journal} {Phys. Rev. B}\ }\textbf {\bibinfo {volume} {88}},\ \bibinfo
  {pages} {085115} (\bibinfo {year} {2013})}\BibitemShut {NoStop}%
\bibitem [{\citenamefont {Barkeshli}\ \emph {et~al.}(2013)\citenamefont
  {Barkeshli}, \citenamefont {Jian},\ and\ \citenamefont
  {Qi}}]{PhysRevB.88.235103}%
  \BibitemOpen
  \bibfield  {author} {\bibinfo {author} {\bibfnamefont {M.}~\bibnamefont
  {Barkeshli}}, \bibinfo {author} {\bibfnamefont {C.-M.}\ \bibnamefont {Jian}},
  \ and\ \bibinfo {author} {\bibfnamefont {X.-L.}\ \bibnamefont {Qi}},\ }\href
  {\doibase 10.1103/PhysRevB.88.235103} {\bibfield  {journal} {\bibinfo
  {journal} {Phys. Rev. B}\ }\textbf {\bibinfo {volume} {88}},\ \bibinfo
  {pages} {235103} (\bibinfo {year} {2013})}\BibitemShut {NoStop}%
\bibitem [{\citenamefont {Santos}\ and\ \citenamefont
  {Hughes}(2017)}]{santos2017}%
  \BibitemOpen
  \bibfield  {author} {\bibinfo {author} {\bibfnamefont {L.~H.}\ \bibnamefont
  {Santos}}\ and\ \bibinfo {author} {\bibfnamefont {T.~L.}\ \bibnamefont
  {Hughes}},\ }\href {\doibase 10.1103/PhysRevLett.118.136801} {\bibfield
  {journal} {\bibinfo  {journal} {Phys. Rev. Lett.}\ }\textbf {\bibinfo
  {volume} {118}},\ \bibinfo {pages} {136801} (\bibinfo {year}
  {2017})}\BibitemShut {NoStop}%
\bibitem [{\citenamefont {Alavirad}\ \emph {et~al.}(2017)\citenamefont
  {Alavirad}, \citenamefont {Clarke}, \citenamefont {Nag},\ and\ \citenamefont
  {Sau}}]{alavirad2017}%
  \BibitemOpen
  \bibfield  {author} {\bibinfo {author} {\bibfnamefont {Y.}~\bibnamefont
  {Alavirad}}, \bibinfo {author} {\bibfnamefont {D.}~\bibnamefont {Clarke}},
  \bibinfo {author} {\bibfnamefont {A.}~\bibnamefont {Nag}}, \ and\ \bibinfo
  {author} {\bibfnamefont {J.~D.}\ \bibnamefont {Sau}},\ }\href {\doibase
  10.1103/PhysRevLett.119.217701} {\bibfield  {journal} {\bibinfo  {journal}
  {Phys. Rev. Lett.}\ }\textbf {\bibinfo {volume} {119}},\ \bibinfo {pages}
  {217701} (\bibinfo {year} {2017})}\BibitemShut {NoStop}%
\bibitem [{\citenamefont {Vaezi}\ and\ \citenamefont
  {Vaezi}(2017)}]{vaezi2017}%
  \BibitemOpen
  \bibfield  {author} {\bibinfo {author} {\bibfnamefont {M.-S.}\ \bibnamefont
  {Vaezi}}\ and\ \bibinfo {author} {\bibfnamefont {A.}~\bibnamefont {Vaezi}},\
  }\href@noop {} {\bibfield  {journal} {\bibinfo  {journal} {arXiv:1706.01192}\
  } (\bibinfo {year} {2017})}\BibitemShut {NoStop}%
\bibitem [{\citenamefont {Wu}\ \emph {et~al.}(2018)\citenamefont {Wu},
  \citenamefont {Wan}, \citenamefont {Kazakov}, \citenamefont {Wang},
  \citenamefont {Simion}, \citenamefont {Liang}, \citenamefont {West},
  \citenamefont {Baldwin}, \citenamefont {Pfeiffer}, \citenamefont
  {Lyanda-Geller},\ and\ \citenamefont {Rokhinson}}]{wu2017}%
  \BibitemOpen
  \bibfield  {author} {\bibinfo {author} {\bibfnamefont {T.}~\bibnamefont
  {Wu}}, \bibinfo {author} {\bibfnamefont {Z.}~\bibnamefont {Wan}}, \bibinfo
  {author} {\bibfnamefont {A.}~\bibnamefont {Kazakov}}, \bibinfo {author}
  {\bibfnamefont {Y.}~\bibnamefont {Wang}}, \bibinfo {author} {\bibfnamefont
  {G.}~\bibnamefont {Simion}}, \bibinfo {author} {\bibfnamefont
  {J.}~\bibnamefont {Liang}}, \bibinfo {author} {\bibfnamefont {K.~W.}\
  \bibnamefont {West}}, \bibinfo {author} {\bibfnamefont {K.}~\bibnamefont
  {Baldwin}}, \bibinfo {author} {\bibfnamefont {L.~N.}\ \bibnamefont
  {Pfeiffer}}, \bibinfo {author} {\bibfnamefont {Y.}~\bibnamefont
  {Lyanda-Geller}}, \ and\ \bibinfo {author} {\bibfnamefont {L.~P.}\
  \bibnamefont {Rokhinson}},\ }\href {\doibase 10.1103/PhysRevB.97.245304}
  {\bibfield  {journal} {\bibinfo  {journal} {Phys. Rev. B}\ }\textbf {\bibinfo
  {volume} {97}},\ \bibinfo {pages} {245304} (\bibinfo {year}
  {2018})}\BibitemShut {NoStop}%
\bibitem [{\citenamefont {Repellin}\ \emph {et~al.}(2018)\citenamefont
  {Repellin}, \citenamefont {Cook}, \citenamefont {Neupert},\ and\
  \citenamefont {Regnault}}]{repellin2017}%
  \BibitemOpen
  \bibfield  {author} {\bibinfo {author} {\bibfnamefont {C.}~\bibnamefont
  {Repellin}}, \bibinfo {author} {\bibfnamefont {A.~M.}\ \bibnamefont {Cook}},
  \bibinfo {author} {\bibfnamefont {T.}~\bibnamefont {Neupert}}, \ and\
  \bibinfo {author} {\bibfnamefont {N.}~\bibnamefont {Regnault}},\ }\href
  {\doibase 10.1038/s41535-018-0085-4} {\bibfield  {journal} {\bibinfo
  {journal} {npj Quantum Materials}\ }\textbf {\bibinfo {volume} {3}},\
  \bibinfo {pages} {14} (\bibinfo {year} {2018})}\BibitemShut {NoStop}%
\bibitem [{\citenamefont {Hutter}\ and\ \citenamefont
  {Loss}(2016)}]{hutter2016}%
  \BibitemOpen
  \bibfield  {author} {\bibinfo {author} {\bibfnamefont {A.}~\bibnamefont
  {Hutter}}\ and\ \bibinfo {author} {\bibfnamefont {D.}~\bibnamefont {Loss}},\
  }\href {\doibase 10.1103/PhysRevB.93.125105} {\bibfield  {journal} {\bibinfo
  {journal} {Phys. Rev. B}\ }\textbf {\bibinfo {volume} {93}},\ \bibinfo
  {pages} {125105} (\bibinfo {year} {2016})}\BibitemShut {NoStop}%
\bibitem [{\citenamefont {Turner}\ \emph {et~al.}(2011)\citenamefont {Turner},
  \citenamefont {Pollmann},\ and\ \citenamefont {Berg}}]{Turner2011}%
  \BibitemOpen
  \bibfield  {author} {\bibinfo {author} {\bibfnamefont {A.~M.}\ \bibnamefont
  {Turner}}, \bibinfo {author} {\bibfnamefont {F.}~\bibnamefont {Pollmann}}, \
  and\ \bibinfo {author} {\bibfnamefont {E.}~\bibnamefont {Berg}},\ }\href
  {\doibase 10.1103/PhysRevB.83.075102} {\bibfield  {journal} {\bibinfo
  {journal} {Phys. Rev. B}\ }\textbf {\bibinfo {volume} {83}},\ \bibinfo
  {pages} {075102} (\bibinfo {year} {2011})}\BibitemShut {NoStop}%
\bibitem [{\citenamefont {Fidkowski}\ and\ \citenamefont
  {Kitaev}(2011)}]{fidkowski2011}%
  \BibitemOpen
  \bibfield  {author} {\bibinfo {author} {\bibfnamefont {L.}~\bibnamefont
  {Fidkowski}}\ and\ \bibinfo {author} {\bibfnamefont {A.}~\bibnamefont
  {Kitaev}},\ }\href {https://link.aps.org/doi/10.1103/PhysRevB.83.075103}
  {\bibfield  {journal} {\bibinfo  {journal} {Phys. Rev. B}\ }\textbf {\bibinfo
  {volume} {83}},\ \bibinfo {pages} {075103} (\bibinfo {year}
  {2011})}\BibitemShut {NoStop}%
\bibitem [{\citenamefont {Bultinck}\ \emph {et~al.}(2017)\citenamefont
  {Bultinck}, \citenamefont {Williamson}, \citenamefont {Haegeman},\ and\
  \citenamefont {Verstraete}}]{Bultinck2017}%
  \BibitemOpen
  \bibfield  {author} {\bibinfo {author} {\bibfnamefont {N.}~\bibnamefont
  {Bultinck}}, \bibinfo {author} {\bibfnamefont {D.~J.}\ \bibnamefont
  {Williamson}}, \bibinfo {author} {\bibfnamefont {J.}~\bibnamefont
  {Haegeman}}, \ and\ \bibinfo {author} {\bibfnamefont {F.}~\bibnamefont
  {Verstraete}},\ }\href {\doibase 10.1103/PhysRevB.95.075108} {\bibfield
  {journal} {\bibinfo  {journal} {Phys. Rev. B}\ }\textbf {\bibinfo {volume}
  {95}},\ \bibinfo {pages} {075108} (\bibinfo {year} {2017})}\BibitemShut
  {NoStop}%
\bibitem [{\citenamefont {Klinovaja}\ and\ \citenamefont
  {Loss}(2014{\natexlab{a}})}]{Klinovaja2014}%
  \BibitemOpen
  \bibfield  {author} {\bibinfo {author} {\bibfnamefont {J.}~\bibnamefont
  {Klinovaja}}\ and\ \bibinfo {author} {\bibfnamefont {D.}~\bibnamefont
  {Loss}},\ }\href {\doibase 10.1103/PhysRevB.90.045118} {\bibfield  {journal}
  {\bibinfo  {journal} {Phys. Rev. B}\ }\textbf {\bibinfo {volume} {90}},\
  \bibinfo {pages} {045118} (\bibinfo {year} {2014}{\natexlab{a}})}\BibitemShut
  {NoStop}%
\bibitem [{\citenamefont {Klinovaja}\ and\ \citenamefont
  {Loss}(2014{\natexlab{b}})}]{Klinovaja2014b}%
  \BibitemOpen
  \bibfield  {author} {\bibinfo {author} {\bibfnamefont {J.}~\bibnamefont
  {Klinovaja}}\ and\ \bibinfo {author} {\bibfnamefont {D.}~\bibnamefont
  {Loss}},\ }\href {\doibase 10.1103/PhysRevLett.112.246403} {\bibfield
  {journal} {\bibinfo  {journal} {Phys. Rev. Lett.}\ }\textbf {\bibinfo
  {volume} {112}},\ \bibinfo {pages} {246403} (\bibinfo {year}
  {2014}{\natexlab{b}})}\BibitemShut {NoStop}%
\bibitem [{\citenamefont {Klinovaja}\ \emph {et~al.}(2014)\citenamefont
  {Klinovaja}, \citenamefont {Yacoby},\ and\ \citenamefont
  {Loss}}]{Klinovaja2014c}%
  \BibitemOpen
  \bibfield  {author} {\bibinfo {author} {\bibfnamefont {J.}~\bibnamefont
  {Klinovaja}}, \bibinfo {author} {\bibfnamefont {A.}~\bibnamefont {Yacoby}}, \
  and\ \bibinfo {author} {\bibfnamefont {D.}~\bibnamefont {Loss}},\ }\href
  {\doibase 10.1103/PhysRevB.90.155447} {\bibfield  {journal} {\bibinfo
  {journal} {Phys. Rev. B}\ }\textbf {\bibinfo {volume} {90}},\ \bibinfo
  {pages} {155447} (\bibinfo {year} {2014})}\BibitemShut {NoStop}%
\bibitem [{\citenamefont {Oreg}\ \emph {et~al.}(2014)\citenamefont {Oreg},
  \citenamefont {Sela},\ and\ \citenamefont {Stern}}]{oreg2014}%
  \BibitemOpen
  \bibfield  {author} {\bibinfo {author} {\bibfnamefont {Y.}~\bibnamefont
  {Oreg}}, \bibinfo {author} {\bibfnamefont {E.}~\bibnamefont {Sela}}, \ and\
  \bibinfo {author} {\bibfnamefont {A.}~\bibnamefont {Stern}},\ }\href
  {\doibase 10.1103/PhysRevB.89.115402} {\bibfield  {journal} {\bibinfo
  {journal} {Phys. Rev. B}\ }\textbf {\bibinfo {volume} {89}},\ \bibinfo
  {pages} {115402} (\bibinfo {year} {2014})}\BibitemShut {NoStop}%
\bibitem [{\citenamefont {Orth}\ \emph {et~al.}(2015)\citenamefont {Orth},
  \citenamefont {Tiwari}, \citenamefont {Meng},\ and\ \citenamefont
  {Schmidt}}]{orth2015}%
  \BibitemOpen
  \bibfield  {author} {\bibinfo {author} {\bibfnamefont {C.~P.}\ \bibnamefont
  {Orth}}, \bibinfo {author} {\bibfnamefont {R.~P.}\ \bibnamefont {Tiwari}},
  \bibinfo {author} {\bibfnamefont {T.}~\bibnamefont {Meng}}, \ and\ \bibinfo
  {author} {\bibfnamefont {T.~L.}\ \bibnamefont {Schmidt}},\ }\href {\doibase
  10.1103/PhysRevB.91.081406} {\bibfield  {journal} {\bibinfo  {journal} {Phys.
  Rev. B}\ }\textbf {\bibinfo {volume} {91}},\ \bibinfo {pages} {081406}
  (\bibinfo {year} {2015})}\BibitemShut {NoStop}%
\bibitem [{\citenamefont {Pedder}\ \emph {et~al.}(2017)\citenamefont {Pedder},
  \citenamefont {Meng}, \citenamefont {Tiwari},\ and\ \citenamefont
  {Schmidt}}]{pedder2017}%
  \BibitemOpen
  \bibfield  {author} {\bibinfo {author} {\bibfnamefont {C.~J.}\ \bibnamefont
  {Pedder}}, \bibinfo {author} {\bibfnamefont {T.}~\bibnamefont {Meng}},
  \bibinfo {author} {\bibfnamefont {R.~P.}\ \bibnamefont {Tiwari}}, \ and\
  \bibinfo {author} {\bibfnamefont {T.~L.}\ \bibnamefont {Schmidt}},\ }\href
  {\doibase 10.1103/PhysRevB.96.165429} {\bibfield  {journal} {\bibinfo
  {journal} {Phys. Rev. B}\ }\textbf {\bibinfo {volume} {96}},\ \bibinfo
  {pages} {165429} (\bibinfo {year} {2017})}\BibitemShut {NoStop}%
\bibitem [{\citenamefont {Vinkler-Aviv}\ \emph {et~al.}(2017)\citenamefont
  {Vinkler-Aviv}, \citenamefont {Brouwer},\ and\ \citenamefont {von
  Oppen}}]{vinkler2017}%
  \BibitemOpen
  \bibfield  {author} {\bibinfo {author} {\bibfnamefont {Y.}~\bibnamefont
  {Vinkler-Aviv}}, \bibinfo {author} {\bibfnamefont {P.~W.}\ \bibnamefont
  {Brouwer}}, \ and\ \bibinfo {author} {\bibfnamefont {F.}~\bibnamefont {von
  Oppen}},\ }\href {\doibase 10.1103/PhysRevB.96.195421} {\bibfield  {journal}
  {\bibinfo  {journal} {Phys. Rev. B}\ }\textbf {\bibinfo {volume} {96}},\
  \bibinfo {pages} {195421} (\bibinfo {year} {2017})}\BibitemShut {NoStop}%
\bibitem [{\citenamefont {Kane}\ and\ \citenamefont {Zhang}(2015)}]{kane2015}%
  \BibitemOpen
  \bibfield  {author} {\bibinfo {author} {\bibfnamefont {C.~L.}\ \bibnamefont
  {Kane}}\ and\ \bibinfo {author} {\bibfnamefont {F.}~\bibnamefont {Zhang}},\
  }\href {http://stacks.iop.org/1402-4896/2015/i=T164/a=014011} {\bibfield
  {journal} {\bibinfo  {journal} {Physica Scripta}\ }\textbf {\bibinfo {volume}
  {T164}},\ \bibinfo {pages} {014011} (\bibinfo {year} {2015})}\BibitemShut
  {NoStop}%
\bibitem [{Note1()}]{Note1}%
  \BibitemOpen
  \bibinfo {note} {See Supplemental Material for more details on bosonization
  and the exact solution to the GS, the construction of non-local parafermions
  and edge Majorana modes and the absence of local parafermions in
  one-dimension fermionic models}\BibitemShut {NoStop}%
\bibitem [{\citenamefont {Zhang}\ and\ \citenamefont {Kane}(2014)}]{zhang2014}%
  \BibitemOpen
  \bibfield  {author} {\bibinfo {author} {\bibfnamefont {F.}~\bibnamefont
  {Zhang}}\ and\ \bibinfo {author} {\bibfnamefont {C.~L.}\ \bibnamefont
  {Kane}},\ }\href {\doibase 10.1103/PhysRevLett.113.036401} {\bibfield
  {journal} {\bibinfo  {journal} {Phys. Rev. Lett.}\ }\textbf {\bibinfo
  {volume} {113}},\ \bibinfo {pages} {036401} (\bibinfo {year}
  {2014})}\BibitemShut {NoStop}%
\bibitem [{\citenamefont {Gogolin}\ \emph {et~al.}(1998)\citenamefont
  {Gogolin}, \citenamefont {Nersesyan},\ and\ \citenamefont
  {Tsvelik}}]{BosonizationGogolin1998}%
  \BibitemOpen
  \bibfield  {author} {\bibinfo {author} {\bibfnamefont {A.}~\bibnamefont
  {Gogolin}}, \bibinfo {author} {\bibfnamefont {A.}~\bibnamefont {Nersesyan}},
  \ and\ \bibinfo {author} {\bibfnamefont {A.}~\bibnamefont {Tsvelik}},\
  }\href@noop {} {\emph {\bibinfo {title} {Bosonization and strongly correlated
  systems}}},\ edited by\ \bibinfo {editor} {\bibfnamefont {C.~U.}\
  \bibnamefont {Press}}\ (\bibinfo {year} {1998})\BibitemShut {NoStop}%
\bibitem [{\citenamefont {Giamarchi}(2004)}]{QP1DGiamarchi2004}%
  \BibitemOpen
  \bibfield  {author} {\bibinfo {author} {\bibfnamefont {T.}~\bibnamefont
  {Giamarchi}},\ }\href@noop {} {\emph {\bibinfo {title} {Quantum Physics in
  One Dimension}}},\ edited by\ \bibinfo {editor} {\bibfnamefont {O.~U.}\
  \bibnamefont {Press}}\ (\bibinfo {year} {2004})\BibitemShut {NoStop}%
\bibitem [{\citenamefont {Iemini}\ \emph {et~al.}(2015)\citenamefont {Iemini},
  \citenamefont {Mazza}, \citenamefont {Rossini}, \citenamefont {Fazio},\ and\
  \citenamefont {Diehl}}]{iemini2015}%
  \BibitemOpen
  \bibfield  {author} {\bibinfo {author} {\bibfnamefont {F.}~\bibnamefont
  {Iemini}}, \bibinfo {author} {\bibfnamefont {L.}~\bibnamefont {Mazza}},
  \bibinfo {author} {\bibfnamefont {D.}~\bibnamefont {Rossini}}, \bibinfo
  {author} {\bibfnamefont {R.}~\bibnamefont {Fazio}}, \ and\ \bibinfo {author}
  {\bibfnamefont {S.}~\bibnamefont {Diehl}},\ }\href {\doibase
  10.1103/PhysRevLett.115.156402} {\bibfield  {journal} {\bibinfo  {journal}
  {Phys. Rev. Lett.}\ }\textbf {\bibinfo {volume} {115}},\ \bibinfo {pages}
  {156402} (\bibinfo {year} {2015})}\BibitemShut {NoStop}%
\bibitem [{\citenamefont {Lang}\ and\ \citenamefont
  {B\"uchler}(2015)}]{lang2015}%
  \BibitemOpen
  \bibfield  {author} {\bibinfo {author} {\bibfnamefont {N.}~\bibnamefont
  {Lang}}\ and\ \bibinfo {author} {\bibfnamefont {H.~P.}\ \bibnamefont
  {B\"uchler}},\ }\href {\doibase 10.1103/PhysRevB.92.041118} {\bibfield
  {journal} {\bibinfo  {journal} {Phys. Rev. B}\ }\textbf {\bibinfo {volume}
  {92}},\ \bibinfo {pages} {041118} (\bibinfo {year} {2015})}\BibitemShut
  {NoStop}%
\bibitem [{\citenamefont {Greiter}\ \emph {et~al.}(2014)\citenamefont
  {Greiter}, \citenamefont {Schnells},\ and\ \citenamefont
  {Thomale}}]{greiter2014}%
  \BibitemOpen
  \bibfield  {author} {\bibinfo {author} {\bibfnamefont {M.}~\bibnamefont
  {Greiter}}, \bibinfo {author} {\bibfnamefont {V.}~\bibnamefont {Schnells}}, \
  and\ \bibinfo {author} {\bibfnamefont {R.}~\bibnamefont {Thomale}},\ }\href
  {\doibase https://doi.org/10.1016/j.aop.2014.08.013} {\bibfield  {journal}
  {\bibinfo  {journal} {Annals of Physics}\ }\textbf {\bibinfo {volume}
  {351}},\ \bibinfo {pages} {1026 } (\bibinfo {year} {2014})}\BibitemShut
  {NoStop}%
\bibitem [{\citenamefont {Katsura}\ \emph {et~al.}(2015)\citenamefont
  {Katsura}, \citenamefont {Schuricht},\ and\ \citenamefont
  {Takahashi}}]{Katsura2015}%
  \BibitemOpen
  \bibfield  {author} {\bibinfo {author} {\bibfnamefont {H.}~\bibnamefont
  {Katsura}}, \bibinfo {author} {\bibfnamefont {D.}~\bibnamefont {Schuricht}},
  \ and\ \bibinfo {author} {\bibfnamefont {M.}~\bibnamefont {Takahashi}},\
  }\href {\doibase 10.1103/PhysRevB.92.115137} {\bibfield  {journal} {\bibinfo
  {journal} {Phys. Rev. B}\ }\textbf {\bibinfo {volume} {92}},\ \bibinfo
  {pages} {115137} (\bibinfo {year} {2015})}\BibitemShut {NoStop}%
\bibitem [{\citenamefont {Bondesan}\ and\ \citenamefont
  {Quella}(2013)}]{bondesan2013}%
  \BibitemOpen
  \bibfield  {author} {\bibinfo {author} {\bibfnamefont {R.}~\bibnamefont
  {Bondesan}}\ and\ \bibinfo {author} {\bibfnamefont {T.}~\bibnamefont
  {Quella}},\ }\href {http://stacks.iop.org/1742-5468/2013/i=10/a=P10024}
  {\bibfield  {journal} {\bibinfo  {journal} {Journal of Statistical Mechanics:
  Theory and Experiment}\ }\textbf {\bibinfo {volume} {2013}},\ \bibinfo
  {pages} {P10024} (\bibinfo {year} {2013})}\BibitemShut {NoStop}%
\bibitem [{Note2()}]{Note2}%
  \BibitemOpen
  \bibinfo {note} {More precisely, the number of fermions is $N_0 + \protect
  \mathaccentV {hat}05EN_e$, where $N_0$ is a reference value.}\BibitemShut
  {Stop}%
\bibitem [{\citenamefont {Haldane}(1981)}]{Haldane1981}%
  \BibitemOpen
  \bibfield  {author} {\bibinfo {author} {\bibfnamefont {F.}~\bibnamefont
  {Haldane}},\ }\href
  {http://iopscience.iop.org/article/10.1088/0022-3719/14/19/010/meta}
  {\bibfield  {journal} {\bibinfo  {journal} {J. Phys. C 14 2585}\ } (\bibinfo
  {year} {1981})}\BibitemShut {NoStop}%
\bibitem [{\citenamefont {Qi}\ \emph {et~al.}(2008)\citenamefont {Qi},
  \citenamefont {Hughes},\ and\ \citenamefont {Zhang}}]{qi2008}%
  \BibitemOpen
  \bibfield  {author} {\bibinfo {author} {\bibfnamefont {X.-L.}\ \bibnamefont
  {Qi}}, \bibinfo {author} {\bibfnamefont {T.~L.}\ \bibnamefont {Hughes}}, \
  and\ \bibinfo {author} {\bibfnamefont {S.-C.}\ \bibnamefont {Zhang}},\
  }\href@noop {} {\bibfield  {journal} {\bibinfo  {journal} {Nat. Physics}\
  }\textbf {\bibinfo {volume} {4}},\ \bibinfo {pages} {273} (\bibinfo {year}
  {2008})}\BibitemShut {NoStop}%
\bibitem [{\citenamefont {Seidel}\ and\ \citenamefont
  {Lee}(2005)}]{seidel2005}%
  \BibitemOpen
  \bibfield  {author} {\bibinfo {author} {\bibfnamefont {A.}~\bibnamefont
  {Seidel}}\ and\ \bibinfo {author} {\bibfnamefont {D.-H.}\ \bibnamefont
  {Lee}},\ }\href {\doibase 10.1103/PhysRevB.71.045113} {\bibfield  {journal}
  {\bibinfo  {journal} {Phys. Rev. B}\ }\textbf {\bibinfo {volume} {71}},\
  \bibinfo {pages} {045113} (\bibinfo {year} {2005})}\BibitemShut {NoStop}%
\bibitem [{Note3()}]{Note3}%
  \BibitemOpen
  \bibinfo {note} {A tight analogy requires the identification $| n,e \rangle +
  | n,o \rangle \to | \theta _n \rangle $ and $| n,e \rangle - | n,o \rangle
  \to | \theta _{n+N/2} \rangle $}\BibitemShut {NoStop}%
\bibitem [{\citenamefont {Fidkowski}\ \emph {et~al.}(2011)\citenamefont
  {Fidkowski}, \citenamefont {Lutchyn}, \citenamefont {Nayak},\ and\
  \citenamefont {Fisher}}]{Fisher2011}%
  \BibitemOpen
  \bibfield  {author} {\bibinfo {author} {\bibfnamefont {L.}~\bibnamefont
  {Fidkowski}}, \bibinfo {author} {\bibfnamefont {R.}~\bibnamefont {Lutchyn}},
  \bibinfo {author} {\bibfnamefont {C.}~\bibnamefont {Nayak}}, \ and\ \bibinfo
  {author} {\bibfnamefont {M.}~\bibnamefont {Fisher}},\ }\href {\doibase
  10.1103/PhysRevB.84.195436} {\bibfield  {journal} {\bibinfo  {journal} {Phys.
  Rev. B 84, 195436}\ } (\bibinfo {year} {2011}),\
  10.1103/PhysRevB.84.195436}\BibitemShut {NoStop}%
\bibitem [{\citenamefont {Cheng}\ and\ \citenamefont {Tu}(2011)}]{cheng2011}%
  \BibitemOpen
  \bibfield  {author} {\bibinfo {author} {\bibfnamefont {M.}~\bibnamefont
  {Cheng}}\ and\ \bibinfo {author} {\bibfnamefont {H.-H.}\ \bibnamefont {Tu}},\
  }\href {\doibase 10.1103/PhysRevB.84.094503} {\bibfield  {journal} {\bibinfo
  {journal} {Phys. Rev. B}\ }\textbf {\bibinfo {volume} {84}},\ \bibinfo
  {pages} {094503} (\bibinfo {year} {2011})}\BibitemShut {NoStop}%
\bibitem [{\citenamefont {Meng}\ \emph {et~al.}(2012)\citenamefont {Meng},
  \citenamefont {Shivamoggi}, \citenamefont {Hughes}, \citenamefont {Gilbert},\
  and\ \citenamefont {Vishveshwara}}]{meng2012}%
  \BibitemOpen
  \bibfield  {author} {\bibinfo {author} {\bibfnamefont {Q.}~\bibnamefont
  {Meng}}, \bibinfo {author} {\bibfnamefont {V.}~\bibnamefont {Shivamoggi}},
  \bibinfo {author} {\bibfnamefont {T.~L.}\ \bibnamefont {Hughes}}, \bibinfo
  {author} {\bibfnamefont {M.~J.}\ \bibnamefont {Gilbert}}, \ and\ \bibinfo
  {author} {\bibfnamefont {S.}~\bibnamefont {Vishveshwara}},\ }\href {\doibase
  10.1103/PhysRevB.86.165110} {\bibfield  {journal} {\bibinfo  {journal} {Phys.
  Rev. B}\ }\textbf {\bibinfo {volume} {86}},\ \bibinfo {pages} {165110}
  (\bibinfo {year} {2012})}\BibitemShut {NoStop}%
\bibitem [{\citenamefont {Jiang}\ \emph {et~al.}(2013)\citenamefont {Jiang},
  \citenamefont {Pekker}, \citenamefont {Alicea}, \citenamefont {Refael},
  \citenamefont {Oreg}, \citenamefont {Brataas},\ and\ \citenamefont {von
  Oppen}}]{liang2013}%
  \BibitemOpen
  \bibfield  {author} {\bibinfo {author} {\bibfnamefont {L.}~\bibnamefont
  {Jiang}}, \bibinfo {author} {\bibfnamefont {D.}~\bibnamefont {Pekker}},
  \bibinfo {author} {\bibfnamefont {J.}~\bibnamefont {Alicea}}, \bibinfo
  {author} {\bibfnamefont {G.}~\bibnamefont {Refael}}, \bibinfo {author}
  {\bibfnamefont {Y.}~\bibnamefont {Oreg}}, \bibinfo {author} {\bibfnamefont
  {A.}~\bibnamefont {Brataas}}, \ and\ \bibinfo {author} {\bibfnamefont
  {F.}~\bibnamefont {von Oppen}},\ }\href {\doibase 10.1103/PhysRevB.87.075438}
  {\bibfield  {journal} {\bibinfo  {journal} {Phys. Rev. B}\ }\textbf {\bibinfo
  {volume} {87}},\ \bibinfo {pages} {075438} (\bibinfo {year}
  {2013})}\BibitemShut {NoStop}%
\bibitem [{\citenamefont {Pientka}\ \emph {et~al.}(2013)\citenamefont
  {Pientka}, \citenamefont {Jiang}, \citenamefont {Pekker}, \citenamefont
  {Alicea}, \citenamefont {Refael}, \citenamefont {Oreg},\ and\ \citenamefont
  {von Oppen}}]{pientka2013}%
  \BibitemOpen
  \bibfield  {author} {\bibinfo {author} {\bibfnamefont {F.}~\bibnamefont
  {Pientka}}, \bibinfo {author} {\bibfnamefont {L.}~\bibnamefont {Jiang}},
  \bibinfo {author} {\bibfnamefont {D.}~\bibnamefont {Pekker}}, \bibinfo
  {author} {\bibfnamefont {J.}~\bibnamefont {Alicea}}, \bibinfo {author}
  {\bibfnamefont {G.}~\bibnamefont {Refael}}, \bibinfo {author} {\bibfnamefont
  {Y.}~\bibnamefont {Oreg}}, \ and\ \bibinfo {author} {\bibfnamefont
  {F.}~\bibnamefont {von Oppen}},\ }\href
  {http://stacks.iop.org/1367-2630/15/i=11/a=115001} {\bibfield  {journal}
  {\bibinfo  {journal} {New Journal of Physics}\ }\textbf {\bibinfo {volume}
  {15}},\ \bibinfo {pages} {115001} (\bibinfo {year} {2013})}\BibitemShut
  {NoStop}%
\bibitem [{\citenamefont {Chew}\ \emph {et~al.}(2018)\citenamefont {Chew},
  \citenamefont {Mross},\ and\ \citenamefont {Alicea}}]{Chew2018}%
  \BibitemOpen
  \bibfield  {author} {\bibinfo {author} {\bibfnamefont {A.}~\bibnamefont
  {Chew}}, \bibinfo {author} {\bibfnamefont {D.~F.}\ \bibnamefont {Mross}}, \
  and\ \bibinfo {author} {\bibfnamefont {J.}~\bibnamefont {Alicea}},\ }\href
  {\doibase 10.1103/PhysRevB.98.085143} {\bibfield  {journal} {\bibinfo
  {journal} {Phys. Rev. B}\ }\textbf {\bibinfo {volume} {98}},\ \bibinfo
  {pages} {085143} (\bibinfo {year} {2018})}\BibitemShut {NoStop}%
\bibitem [{\citenamefont {Calzona}\ \emph {et~al.}(2018)\citenamefont
  {Calzona}, \citenamefont {Meng}, \citenamefont {Sassetti},\ and\
  \citenamefont {Schmidt}}]{Calzona2018}%
  \BibitemOpen
  \bibfield  {author} {\bibinfo {author} {\bibfnamefont {A.}~\bibnamefont
  {Calzona}}, \bibinfo {author} {\bibfnamefont {T.}~\bibnamefont {Meng}},
  \bibinfo {author} {\bibfnamefont {M.}~\bibnamefont {Sassetti}}, \ and\
  \bibinfo {author} {\bibfnamefont {T.~L.}\ \bibnamefont {Schmidt}},\
  }\href@noop {} {\bibfield  {journal} {\bibinfo  {journal} {arXiv:1802.06061}\
  } (\bibinfo {year} {2018})}\BibitemShut {NoStop}%
\bibitem [{Note4()}]{Note4}%
  \BibitemOpen
  \bibinfo {note} {In the spin language, $\protect \mathaccentV {tilde}07Ec_j$
  corresponds to the operator $S_j^-$ flipping the local spin from up to
  down.}\BibitemShut {Stop}%
\bibitem [{\citenamefont {Prosen}(2008)}]{Prosen2008}%
  \BibitemOpen
  \bibfield  {author} {\bibinfo {author} {\bibfnamefont {T.}~\bibnamefont
  {Prosen}},\ }\href@noop {} {\bibfield  {journal} {\bibinfo  {journal} {New
  Journal of Physics}\ }\textbf {\bibinfo {volume} {10}},\ \bibinfo {pages}
  {043026} (\bibinfo {year} {2008})}\BibitemShut {NoStop}%
\bibitem [{\citenamefont {Jaffe}\ and\ \citenamefont
  {Pedrocchi}(2015)}]{Jaffe2015}%
  \BibitemOpen
  \bibfield  {author} {\bibinfo {author} {\bibfnamefont {A.}~\bibnamefont
  {Jaffe}}\ and\ \bibinfo {author} {\bibfnamefont {F.~L.}\ \bibnamefont
  {Pedrocchi}},\ }\href@noop {} {\bibfield  {journal} {\bibinfo  {journal}
  {Communications in Mathematical Physics}\ }\textbf {\bibinfo {volume}
  {337}},\ \bibinfo {pages} {455} (\bibinfo {year} {2015})}\BibitemShut
  {NoStop}%
\end{thebibliography}%

\end{document}